\documentclass[a4paper,11pt]{article}
\pdfoutput=1 

\usepackage{jinstpub} 

\usepackage{lineno}
\usepackage{rotating}

\title{\boldmath An 8 GeV linac as the Booster replacement in the Fermilab Power Upgrade}

\author[a]{D. Neuffer,\note{Corresponding author.}}
\author[a,b,1]{S. Belomestnykh,}
\author[a]{D. Johnson,}
\author[a,c]{H. Padamsee,}
\author[a]{S. Posen,}
\author[a]{E.~Pozdeyev,}
\author[a]{V. Pronskikh,}
\author[a]{A. Saini,}
\author[a]{N. Solyak,}
\author[a]{V. Yakovlev}

\affiliation[a]{Fermi National Accelerator Laboratory,\\Batavia, IL, USA}
\affiliation[b]{Stony Brook University,\\Stony Brook, NY, USA}
\affiliation[c]{Cornell University,\\Ithaca, NY, USA}

\emailAdd{neuffer@fnal.gov}

\abstract{Increasing the Main Injector beam power above $\sim$1.2 MW requires replacement of the 8 GeV Booster by a higher intensity alternative. In this paper, we consider an 8 GeV linac Booster replacement that produces 8 GeV H$^-$ beam for injection into the Recycler Ring or Main Injector.  This upgrade will maximize the beam available for neutrino production for the long baseline DUNE experiment to greater than 2.5 MW and enable a next generation of intensity frontier experiments. The 8 GeV linac takes $\sim$1 GeV beam from the PIP-II Linac and accelerates it to $\sim$2 GeV in a 650~MHz superconducting RF linac, followed by a $\sim$2 to 8 GeV pulsed linac using 1300 MHz cryomodules. The linac components incorporate recent improvements in superconducting RF technology. The linac configuration and beam dynamics requirements are presented. Injection options are discussed, including use of an 8~GeV Accumulator Ring. Foil-based injection is the present standard but R\&D toward implementing laser-assisted injection could enable a significant improvement. Research needed to implement the Booster replacement is described.}


\arxivnumber{2203.05052} 

\begin{document}
\maketitle
\flushbottom

\section{Introduction}

The Fermilab Proton Improvement Plan II (PIP-II) project will provide an 800 MeV proton beam with CW capability, with beam power up to the MW level available for user experiments \cite{1}. However, the amount of beam that can be transmitted to the Main Injector (MI) is limited by the existing 0.8 – 8.0 GeV Booster ring capacity. The next Fermilab upgrade should include a replacement for the Booster. The Project X design proposal included some options for that replacement, based on a continuation of the 800 MeV linac to 2 – 3 GeV followed by either a Rapid Cycling Synchrotron (RCS) or continuing the linac to 8 GeV \cite{2}. An 8 GeV linac may be made relatively affordable by extending the use of 650 MHz PIP-II cryomodules and adding relatively inexpensive  1300 MHz ILC-style superconducting RF (SRF) cryomodules, that have already been designed and mass-produced for the European XFEL and LCLS-II machines.

In this paper we focus on the 8 GeV Booster Replacement Linac (BRL). We begin with discussion of the beam requirements and potential layouts for the linac after PIP-II in sections 2, 3, and 4. The Project X 8 GeV design is used as an initial template. Constraints on accelerating gradients and magnetic fields are presented in sections 5 and 6. Sections 7 and 8 address injection into MI and/or Recycler Ring (RR). The BRL AC power estimate is presented in section~9. In section~10 we discuss three scenarios for operation of the 2 -- 8 GeV linac. Finally, section 11 outlines R\&D necessary to implement the Booster replacement. We conclude the paper with a brief summary.

\section{ Linac scenario requirements}

The Fermilab PIP-II project will provide a new 800 MeV SRF linac that replaces the existing 400~MeV normal conducting linac, enabling higher intensity injection into the Fermilab Booster and providing 800 MeV proton beam to other experiments. The primary purpose of PIP-II is to provide enhanced beam power delivery from the Main Injector to Deep Underground Neutrino Experiment (DUNE). This is enabled by increasing the beam energy and intensity delivered by the linac to the Booster and increasing the Booster cycle rate, thereby increasing injected beam into the MI. Table~\ref{tab:i} shows high-level parameters of the proton beam to DUNE before PIP-II (PIP) and after PIP-II, as presented in the Fermilab PIP-II Design Report. PIP-II increases the Booster cycle rate to 20 Hz and the beam intensity to $6.5\times10^{12}$ protons/pulse, enabling MI beam power of $\sim$1 – 1.2 MW at beam energies of 60 to 120 GeV.

\begin{table}[htbp]
\centering
\caption{\label{tab:i} High level performance goals for PIP, PIP-II, and Booster Replacement Linac (BRL).}
\smallskip
\begin{tabular}{|l|l|l|l|l|}
\hline
\textbf{Performance Parameter} & \textbf{PIP} & \textbf{PIP-II} & \textbf{BRL} & \textbf{Units}\\
\hline
Linac beam energy & 400 & 800 & 8000 & MeV\\
Linac beam current (chopped) & 25 & 2 & 2 & mA\\
Linac pulse length & 0.03 & 0.54 & 2.2 & ms\\
Linac pulse repetition rate & 15 & 20 & 20 & Hz\\
Linac upgrade potential & N/A & CW & CW & \\
8 GeV protons per pulse (extracted) & 4.2 & 6.5 & 27.5 & $10^{12}$\\
8 GeV pulse repetition rate & 15 & 20 & 20 & Hz\\
Beam power at 8 GeV & 80 & 166 & 700 & kW\\
8 GeV beam power to MI & 50 & 83 -- 142* & 176 -- 300 & kW\\
Beam power to 8 GeV other (pulsed mode) & 30 & 83 -- 24* & 500 -- 375 & kW\\
Main Injector protons per Pulse (extracted) & 4.9 & 7.5 & 15.6 & $10^{13}$\\
Main Injector cycle time at 120 GeV & 1.33 & 1.2 & 1.2 & s\\
Main Injector cycle time at 60 GeV & N/A & 0.7 & 0.7 & s\\
Beam power at 120 GeV & 0.7 & 1.2 & 2.5 & MW\\
Beam power at 60 GeV & N/A & 1 & 2.15 & MW\\
\hline
\multicolumn{5}{l}{\small *Total PIP-II with Booster 8 GeV power is 166 kW.} \\
\end{tabular}
\end{table}

With the completion of PIP-II, the 60-years-old Booster becomes the limiting bottleneck in providing MI beam to DUNE, and further improvements will require replacement of the Booster with a higher-capacity injector. This Booster Replacement (BR) should provide substantially higher intensity to DUNE. The initial design specification for the BR upgrade is that it should enable at least 2.4 MW from the MI. High-level performance goals for a Booster Replacement Linac are presented in Table~\ref{tab:i}. The BRL parameters presented there should be considered as initial goals for the BRL project. The BRL project should be capable of significant extension beyond these initial goals, and these potential improvements may influence the BRL design development.

Booster replacement was considered by the Project X research program and both linac and RCS versions were developed \cite{2}. The parameters for the BRL shown in Table~\ref{tab:i} are based in part on the Project X parameters, updated to follow recently developed requirements.

\begin{figure}[htbp]
\centering 
\includegraphics[width=.9\textwidth]{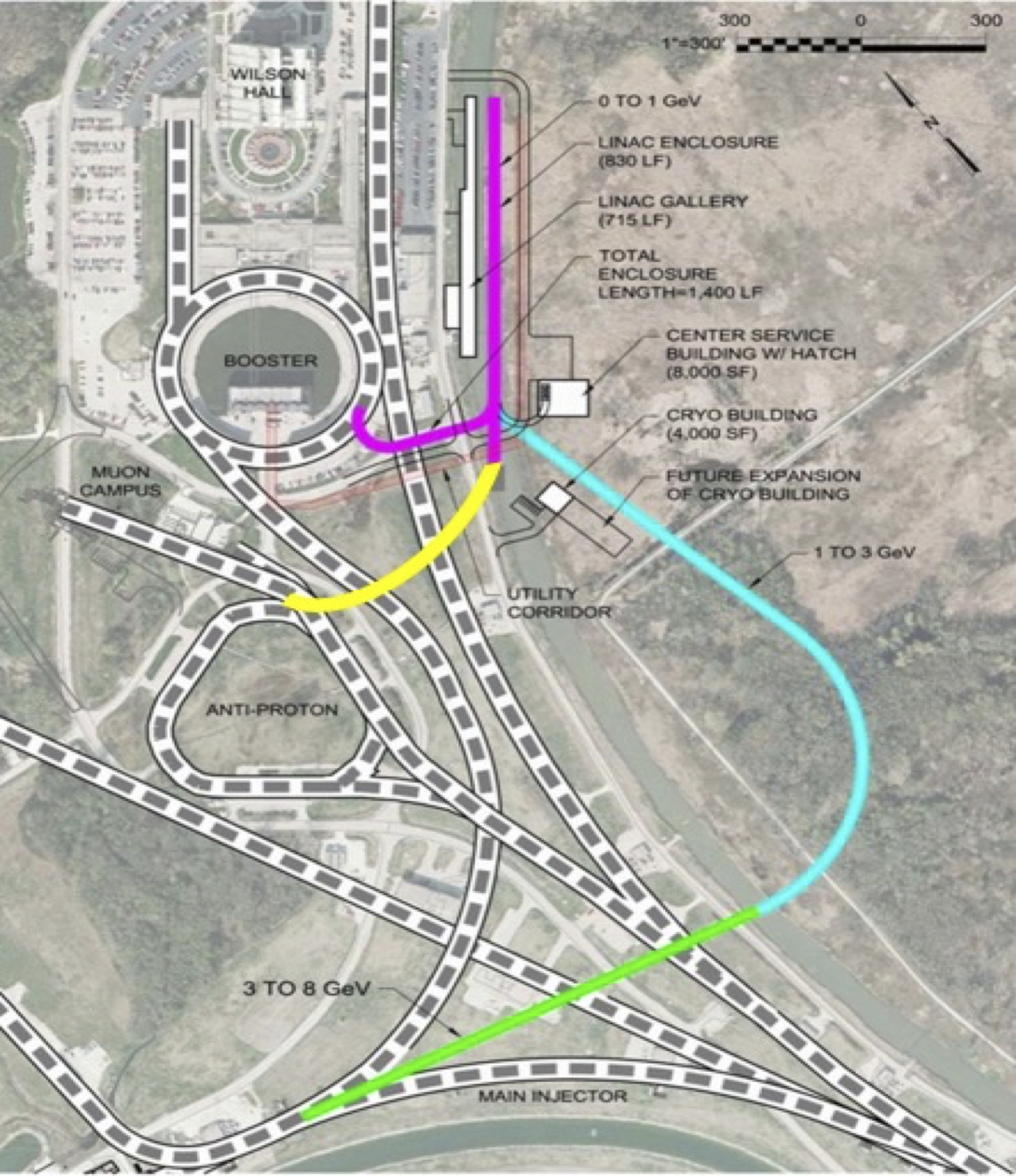}
\caption{\label{fig:1} Layout of the 8 GeV linac as envisioned in Project X.}
\end{figure}

Fig.~\ref{fig:1} shows a potential layout for the BRL, based upon the Project X design \cite{3}. The 800~MeV linac is extended to $\sim$1 GeV. The beam exiting that linac is bent away from the Main Injector by $\sim45^\circ$ into a 1 to 3~GeV linac ($\sim$280 m long), consisting of 650 MHz cryomodules. This is followed by a $\sim105^\circ$ bend, that turns the beam back toward the Main Injector. A $\sim$390 m long linac, consisting of 1300 MHz cryomodules accelerates the beam from 3 to 8 GeV, and a following short transport injects the H$^-$ beam into the Recycler Ring directly above the Main Injector, where charge exchange injection accumulates protons, for transfer into the Main Injector. Parameters of the linac components are shown in Table~\ref{tab:ii}. The curves away and back toward the Main Injector are required to fit the relatively long linac sections into the short space between the PIP-II linac and the Main Injector.
This Project X scenario was considered as an initial guide toward constructing the 8 GeV linac version presented in this report.

\begin{table}[htbp]
\centering
\caption{\label{tab:ii} Parameters of the Project X 8 GeV linac.}
\smallskip
\begin{tabular}{|l|l|l|l|l|l|}
\hline
\textbf{Section} & \textbf{Length} & \textbf{Bending field} & \textbf{Bending} & \textbf{Cavities/} & \textbf{Cryomodule}\\
 & & \textbf{or} & \textbf{angle or} & \textbf{magnets/} & \textbf{length} \\
& & \textbf{RF frequency} & \textbf{Linac mode} & \textbf{cryomodules} & \\
\hline
1 GeV transport & 40 m & 0.277 T & $-45^\circ$ & & \\
1 to 3 GeV linac & 240 m & 650 MHz & CW & 120/20/20 & 9.92 m \\
3 GeV bend & 200 m & 0.13 T & $105^\circ$ & & \\
3 to 8 GeV linac & 390 m & 1300 MHz & Pulsed, 10 Hz& 224/28/28 & 12.5 m \\
8 GeV injection & & 0.055 T & & & \\
\hline
\end{tabular}
\end{table}

\subsection{Advantages and disadvantages of a linac upgrade}

We are considering both linac and rapid-cycling synchrotron (RCS) versions of the BR. The RCS version is presented in some detail in \cite{4}.

\paragraph{\textit{Potential advantages of the linac option include:}}
\begin{itemize}
    \item Improvements in SRF technology have greatly reduced the cost of accelerating cavities and cryomodules. 650 MHz cryomodules will be developed for PIP-II and the incremental construction costs of the additional modules of an 8 GeV linac scenario should be affordable. 1300~MHz cryomodules are being mass-produced for electron accelerators. The same cryomodules can be used for high-energy proton acceleration, and these should also be affordable.
    \item Linac acceleration through to the MI eliminates the need for intermediate injections and accumulator rings, which could simplify construction and operation.
    \item It should be substantially more efficient than an RCS. An RCS system expends considerable energy in cycling magnets from low to high field and back, in addition to the accelerating RF power requirements. SRF cavities are very efficient in converting RF power into beam energy. This efficiency advantage would be magnified in a higher-power upgrade to CW operation. The linac system could be upgradable to a much larger total delivered beam power.
    \item Many experiments benefit from a CW beam, which is most readily obtained from a CW linac.
    \item While a foil-stripping injection may be more difficult at 8 GeV, laser-assisted injection may be easier because longer-wavelength lasers can be used.
\end{itemize}
	 
\paragraph{\textit{Potential disadvantages of the linac option include:}}	
\begin{itemize}
    \item Requires H$^-$ injection at 8 GeV into the Main Injector or Recycler Ring. The MI/RR is not designed to include charge exchange injection; modifications are needed.
    \item H$^-$ beam at 8 GeV is vulnerable to magnetic stripping, gas stripping and intra-beam stripping.
    \item The PIP-II to MI geometry is constrained and relatively inflexible. The most natural injection point would be MI-10, which is slated for Long Baseline Neutrino Facility (LBNF) extraction. The alternative is RR-10 injection with beam accumulation in the RR, with transfer to the MI. The injection difficulties are discussed in more detail below.
    \item Many experiments need pulsed beam. This can be obtained by H$^-$ injection and accumulation into a fixed energy storage ring, at the cost of an added $\sim$8 GeV storage ring. This storage ring could also be used to accumulate beam for injection into the MI, providing an alternative to RR-10 injection.
    \item H$^-$ injection into a new RCS ring could be more optimized, since a new ring could have straight sections more completely optimized for foil injection. Also, losses would occur with lower energy injected beam rather than 8 GeV linac injected beam.
\end{itemize}

\section{PIP-II linac design and parameters}

The Booster Replacement Linac would be an extension of the PIP-II linac \cite{1}. The PIP-II project encompasses a set of upgrades and improvements to the Fermilab accelerator complex aimed at supporting a world-leading High Energy Physics program over the next several decades. The primary goals for PIP-II are:
\begin{itemize}
    \item Deliver beam with a power of 1.2 MW to the LBNF/DUNE target, upgradable to multi-MW.
    \item Deliver a platform capable of high-duty-factor/high-beam-power operations while providing flexible bunch patterns to multiple experiments simultaneously.
    \item Deliver a platform to support future upgrades of the accelerator complex.
    \item Ensure sustained high reliability of the Fermilab accelerator complex.
    \item The above capabilities should be provided in a cost-effective manner.
\end{itemize}   

\begin{figure}[htbp]
\centering 
\includegraphics[width=1.0\textwidth]{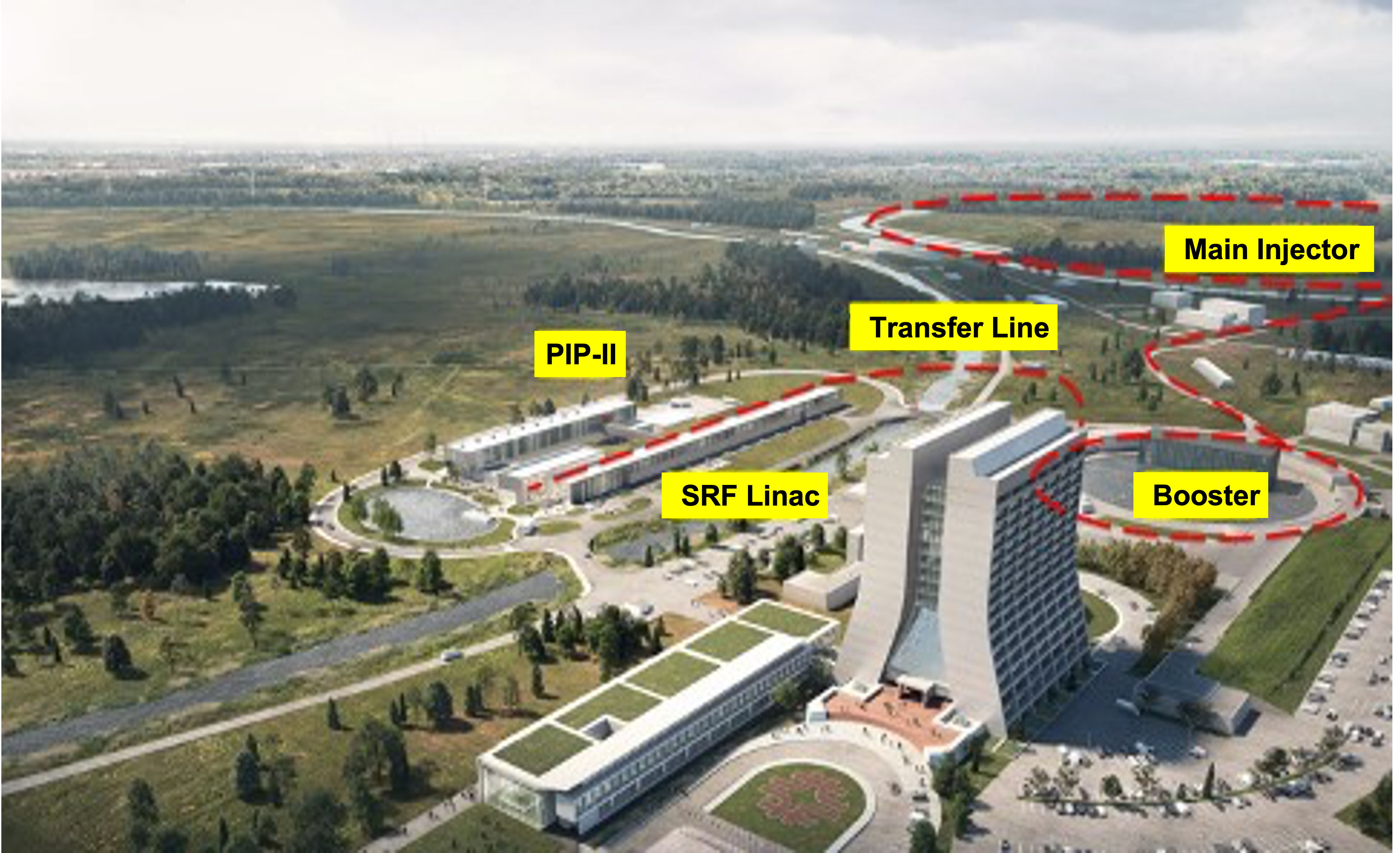}
\caption{\label{fig:2} PIP-II linac location on Fermilab campus.}
\end{figure}

\begin{figure}[htbp]
\centering 
\includegraphics[width=1.0\textwidth]{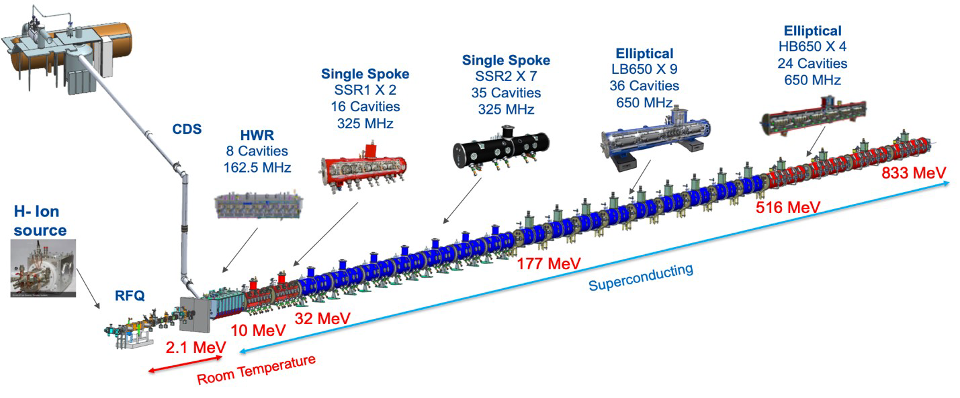}
\qquad
\caption{\label{fig:3} PIP-II Linac with the cryoplant and the cryogenic distribution system (CDS). The linac consists of the room-temperature front end, one half-wavelength resonator (HWR) cryomodule, two types of single-spoke resonator cryomodules (SSR1 and SSR2), and two types of elliptical cavity cryomodules (LB650 and HB650).}
\end{figure}

PIP-II includes a superconducting linac to fuel the next generation of intensity frontier experiments. The linac will accelerate H$^-$ ions to 800 MeV for injection into the Booster. The project also includes upgrades to the existing Booster, Main Injector, and Recycler rings that will enable them to operate at an increased repetition rate (Booster at 20 Hz and Main Injector at 0.83 Hz) and deliver a 1.2 MW proton beam to the LBNF target. Fig.~\ref{fig:2} shows the PIP-II linac location with the beam transfer line to the Booster on the Fermilab campus. Fig.~\ref{fig:3} shows the layout of the major linac components.

\paragraph{\textit{The PIP-II linac components include:}}
\begin{itemize}
    \item \textit{Ion sources and LEBT.} The baseline design of the PIP-II Linac includes two identical multi-cusp, filament-driven, H$^-$ sources with their own low energy beam transport (LEBT) branches. A 3-way switching magnet allows fast switching between the sources. The ion sources are designed to operate in the DC regime, producing up to approximately 15 mA of H$^-$ beam current.
    \item \textit{RFQ.} The PIP-II RFQ is a 4-vane brazed structure operating at 162.5 MHz. The RFQ is designed to accelerate the beam to 2.1 MeV in the CW regime. The RFQ sets the minimum temporal separation between bunches in the linac.
    \item \textit{MEBT Chopper.} The medium energy beam transport (MEBT) includes a chopper that is designed to selectively remove bunches without affecting neighboring bunches. The chopper can create arbitrarily programmed bunch patterns. The chopper consists of two kickers, a beam absorber, and pulse forming electronics. The beam intensity is reduced by approximately 60\% in the MEBT by the fast MEBT chopper and collimators to ensure lossless injection of the 162.5 MHz PIP-II bunch pattern into the 45 MHz RF of the Booster.
    \item \textit{SRF Linac.} The SRF linac is designed to accelerate H$^-$ beams and consists of five different types of cavities as show in Fig.~\ref{fig:3}. The number of cavities and their design are optimized to match the velocity profile of the accelerated H$^-$ beam. The scope of the PIP-II project includes four HB650 cryomodules, sufficient to accelerate the beam to 833 MeV. The linac tunnel has space for two additional HB650 CMs that can further accelerate the beam to 1050~MeV. The design of the RF cavities is optimized for CW operations with a beam intensity of several milliamperes.
    \item \textit{Beam Transfer Line (BTL).} The PIP-II project includes a beam transfer line to deliver the beam from the linac to the Fermilab Booster.
\end{itemize}

\paragraph{\textit{The initial PIP-II design establishes some key features:}}
\begin{itemize}
    \item The MEBT chopper can provide arbitrary, programmable bunch patterns, including gaps and reduced frequency. This functionality is critical for PIP-II operations. Because the frequency of the linac RF is not a harmonic of the Booster RF or the MI RF, a significant portion of the linac bunches will be lost. The chopper selectively removes bunches that would be injected too close to the separatrix or miss the bucket. Based on simulations of the injection process into the Booster, the chopper will have to remove up to 60\% of the bunches. In addition, the chopper will be used to produce flexible bunch patterns required for other users and secondary beam operation.
    \item The highest bunch frequency is 162.5 MHz, determined by the RFQ. The bunch frequency can be reduced using the MEBT chopper.
    \item The number of H$^-$ per bunch can reach $4\times10^8$ without suffering significant degradation of the beam quality. The nominal bunch intensity in the LBNF mode is $1.9\times10^8$ H$^-$ per bunch.
    \item The average beam current is limited to 2 mA ($1.25\times10^{16}$ H$^-$/sec) by the RF power available from the RF amplifiers. Any combination of bunch frequency and charge is possible if the average current in the pulse does not exceed 2 mA, the bunch frequency does not exceed 162.5 MHz, and the maximum number of particles per bunch does not exceed $4\times10^8$. 
\end{itemize}

\subsection{PIP-II science program with beam provided to multiple users}

The PIP-II Mission Need Statement (MNS) requires PIP-II to deliver 1.2 MW of beam power from the MI onto the LBNF target, and to provide beam to other laboratory experiments. The MNS also requires PIP-II to allow a subsequent doubling of beam power delivered from the MI while providing beam to other users for a broader spectrum of particle physics research opportunities. The design of the PIP-II linac includes provisions that facilitate future upgrades and addition of users. The Booster Replacement program should also provide further opportunities for other users, with the possibility to provide proton beam to new experiments using 1 – 8 GeV protons. 

In addition to the MI program, the PIP-II linac (with no energy upgrade) can deliver a CW beam with a power of up to 1.6 MW. The DUNE experiment in the PIP-II era (1.2 MW on the LBNF target) requires only approximately 1.1\% of the total potential beam intensity. Even with the doubled beam power on target, the LBNF beam will require only 2.2\% of the linac CW intensity. The rest of the beam can be delivered to multiple users, enabling concurrent operations.

\begin{figure}[htbp]
\centering 
\includegraphics[width=1.0\textwidth]{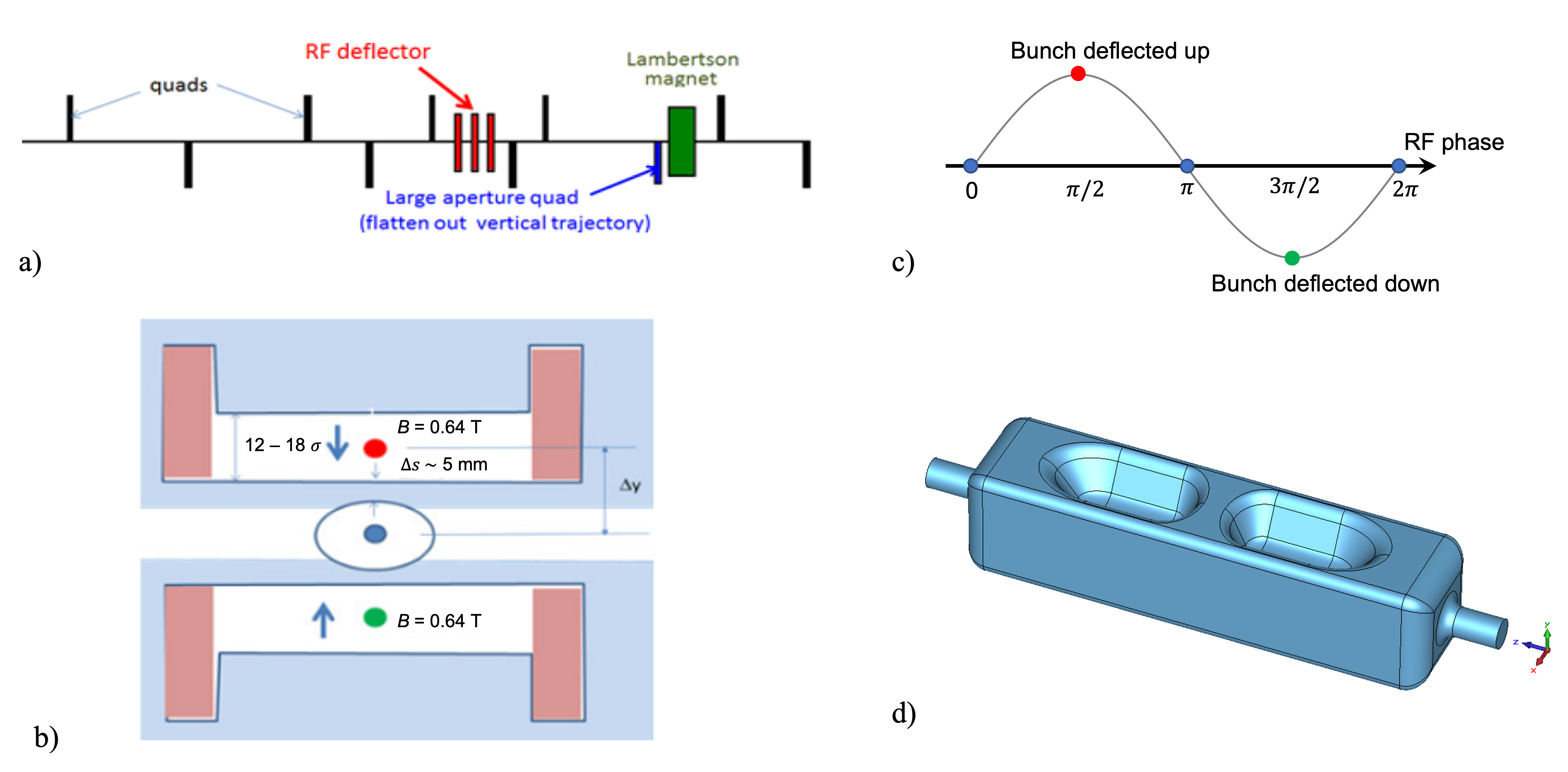}
\qquad
\caption{\label{fig:4} a) RF separator concept. b) 3-way Lambertson magnet. c) The amplitude and the direction of the deflection angle depend on the phase of bunch arrival in the cavity. d) A 3D model of SRF cavity designed to provide a transverse kick, operating at a subharmonic of the bunch repetition frequency.}
\end{figure}

There are two main types of devices that can be used to distribute beam to multiple users:
\begin{itemize}
    \item \textit{RF separators} are based on RF cavities that operate at a subharmonic of the bunch repetition frequency and separate the beam into two or more beamlets. The separation scheme \cite{39} is illustrated in Fig.~\ref{fig:4}a. It contains RF deflecting cavities which separate the beam in the vertical plane to select an aperture of a horizontal bending Lambertson magnet (Fig.~\ref{fig:4}b). Instead of accelerating the beam, RF separators are designed to provide a transverse kick using either electric or magnetic RF fields. Fig.~\ref{fig:4}c shows the principle of operation of an RF deflector and Fig.~\ref{fig:4}d depicts a 3D model of SRF deflecting cavity. The kick provided by the separator depends on the bunch arrival phase.
    \item \textit{Fast-switching magnets} can deflect the beam to required experiments and switch between beam destinations in 10 – 20 microseconds. Such a magnet can be programmed to switch the beam periodically between multiple users in a quasi-concurrent manner, delivering periodic bursts of beam with the full pulse intensity. The fast MEBT chopper will turn off the beam in the linac during magnet switching to avoid beam losses.
\end{itemize}

\subsection{Intensity upgrade}

To achieve 1.2 MW of beam power on the LBNF target, PIP-II needs to accelerate 0.54-ms-long beam pulses with an average peak current of 2 mA. This peak beam intensity is an order of magnitude lower than that in some other pulsed, high-power accelerators such as SNS or ESS. This choice provides several advantages. First, it reduces intensity-dependent effects and alleviates their impact on the beam quality, simplifying beam chopping and allowing for precision painting during injection in the Booster. Second, it allows using widely available, easy-to-operate, CW-capable solid-state amplifiers. On the other hand, the low beam intensity requires longer injection time into the Booster and will require longer injection times for the 8 GeV linac into the MI. Long injection times can be problematic due to injection foil overheating and cycle timing constraints. Increasing the beam current in the linac can mitigate these issues. 

In considering options for boosting the linac intensity, we divide the intensity increase into two ranges loosely based on the impact on accelerator systems and beam parameters: a moderate increase by roughly a factor of 2 and a substantial increase by a factor of 5 to 10:

\textbf{Moderate increase of linac beam current.} Results of numerical studies and engineering estimates show that increasing the beam current to 4 – 5 mA, that is, by a factor of 2 to 2.5 relative to the PIP-II baseline design, is feasible without significant design changes and requires only increasing the output power of RF amplifiers roughly proportionally to the beam current. The increased RF power output can affect requirements for the facility electrical power, utilities, and space. Therefore, more compact and efficient amplifier designs shall be evaluated as an alternative to solid state amplifiers.

\textbf{Substantial increase of linac beam current.} An increase of the linac beam current by a factor of 5 to 10 (from 2 mA to 10 mA to 20 mA) will require more substantial changes to the PIP-II facility. A new ion source and RFQ would be required.  Space charge and intra-beam scattering will become important limitations in operation. RF amplifiers, power supplies and RF power couplers will need to be upgraded. The linac will require significant modifications as well. In the present study we limit considerations to moderate beam current increase only.

\section{BRL scenario description}

\begin{figure}[htbp]
\centering 
\includegraphics[width=.9\textwidth]{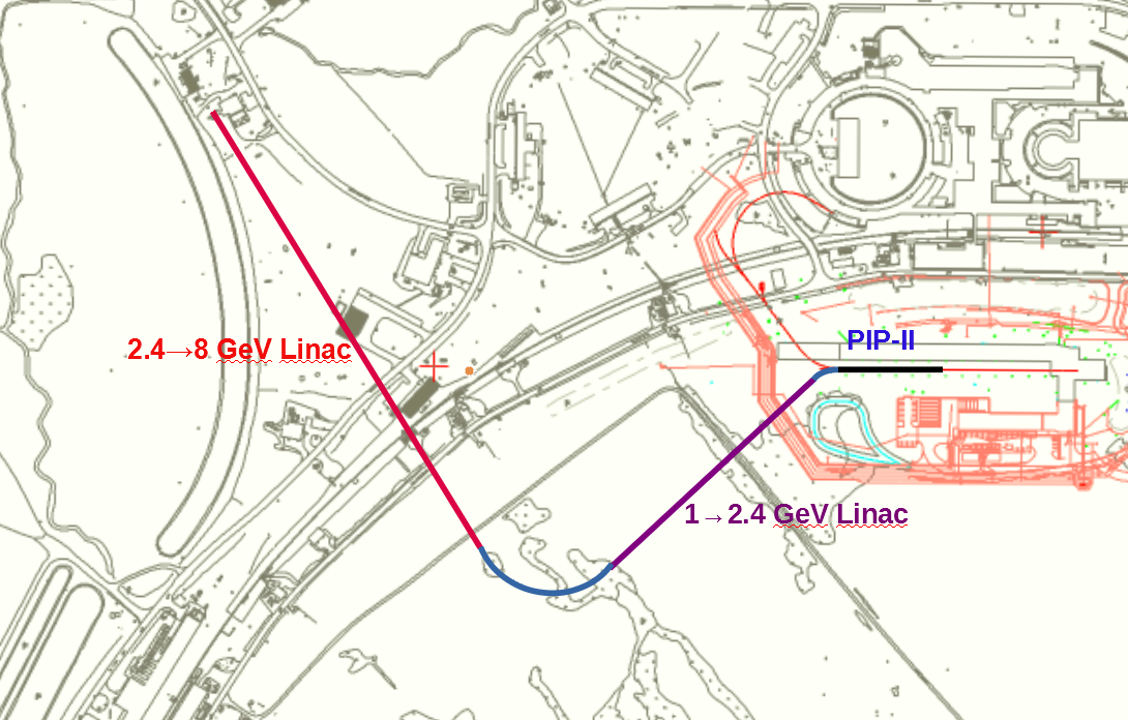}
\qquad
\caption{\label{fig:5} Layout of the 8 GeV linac of Table~\ref{tab:iii} superimposed on a recent PIP-II layout map; note similarities to Fig.~\ref{fig:1}.}
\end{figure}

The PIP-II linac provides the basis for the Booster Replacement Linac scenarios. In this section we describe a baseline approach for the BRL, starting from the currently established PIP-II location. Fig.~\ref{fig:5} shows the layout, which is similar to one considered in the Project X era, and Table~\ref{tab:iii} summarizes the component parameters. The initial PIP-II linac is extended to 1 GeV extraction, by inserting 2 additional cryomodules in the existing space at the end of the linac.  A switching magnet at that point bends the beam into a curved channel (away from the Booster toward Fermilab South). This curved channel bends the beam by $\sim45^\circ$, leading into a 650 MHz linac which takes the beam to 2.4 GeV. This is followed by a $\sim105^\circ$ curved transport that directs the beam into a 1300 MHz linac that accelerates the beam to 8 GeV. Finally, a transport line matches the H$^-$ beam into the Recycler at RR-10.

\begin{table}[htbp]
\centering
\caption{\label{tab:iii} Components of the 8 GeV linac scenario, starting from 1 GeV PIP-II extraction. The geometry is similar to the Project X case.}
\smallskip
\begin{tabular}{|l|l|l|l|l|l|}
\hline
\textbf{Section} & \textbf{Length} & \textbf{Bending field} & \textbf{Total bending} & \textbf{Cavities/} & \textbf{Cryomodule}\\
 & & \textbf{or} & \textbf{angle or} & \textbf{magnets/} & \textbf{length} \\
& & \textbf{RF frequency} & \textbf{Linac mode} & \textbf{cryomodules} & \\
\hline
1 GeV transport & 32 m & 0.25 T & $-45^\circ$ & & \\
1 to 2.4 GeV linac & 290 m & 650 MHz & CW or 20 Hz & 66/11/11 & 9.92 m \\
2.4 GeV bend & 165 m & 0.13 T & $105^\circ$ & & \\
2.4 to 8 GeV linac & 310 m & 1300 MHz & Pulsed, 20 Hz& 160/20/20 & 12.5 m \\
8 GeV injection & 200 m & 0.05 T & & & \\
\hline
\end{tabular}
\end{table}

In our initial configuration, the beam pulse structure is similar to that for the PIP-II Booster injection.  The beam current is pulsed at 20 Hz, with each pulse 2.2 ms long and the beam current is chopped from 5 mA to 2 mA to match the ~53 MHz pattern of beam injected into the MI/RR. 6~pulses of the H$^-$ beam are used to fill the RR in preparation for transfer to the Main Injector and acceleration. The MI cycle is designed to cycle in 1.2 s to 120 GeV for extraction toward the LBNF target, resulting in $\sim$2.5 MW power of beam on target for the DUNE neutrino beam. An alternative cycle of 0.7 s at 60 GeV (2.2 MW) will also be possible.

In this initial pulsed configuration, only $1/4$ of the 20 Hz pulses are directed toward the RR/MI, and the remaining pulses could be directed toward other experiments using 8 GeV proton beam, or beam could be extracted at intermediate energies, such as 2.4 GeV. The PIP-II linac is capable of CW operation, and the BRL 650 and 1300 MHz linacs could also be upgraded to CW, if the physics program demands it.

\subsection{Transport from PIP-II}

The Transport from PIP-II is initiated by a fast-switching magnet that deflects the beam into a transport section that bends the beam trajectory by $45^\circ$ and directs it into the first linac section. The transport would be an achromat of four $90^\circ$ cells, similar to that used for transport into the Booster.

\subsection{650 MHz linac}

In the geometry of Fig.~\ref{fig:5}, the linac section between the arcs is $\sim$290 m long, and it is designed to accelerate the beam to an intermediate energy before it is directed into the final linac section.  The value of that intermediate energy will depend upon demand for physics at that energy as well as cost and performance optimization. Another consideration is to keep the energy low enough to avoid magnetic stripping in the $105^\circ$ bend toward the MI. Values of 2 to 3 GeV for this transition have been considered, with the higher value limited by the magnetic stripping and the lower value by cost optimization and the threshold for kaon production. In our initial scenario we have chosen 2.4 GeV as the reference case. 

The linac could be composed of 650 MHz or 1300 MHz SRF cryomodules, with the transition occurring at an optimal energy. Previous studies conducted to compare the efficiency of HB650 and TESLA 1.3 GHz cryomodules for the linac extension concluded that 650 MHz HB650 cavities and cryomodules were more efficient below 2 GeV, and possibly as high as 3 GeV. By the time the PIP-II energy upgrade will be ready to be implemented, the design and performance of the HB650 cryomodule and its cavities will be validated and significant experience with their manufacturing, testing, and operation will be obtained, so that a more precise optimum could be determined. Recent advances in the 1300 MHz SRF technology for CW operation in LCLS-II will be taken into account as well. To simplify the initial configuration of the BRL, the 650 MHz PIP-II cryomodules are chosen for this entire segment, see Fig.~\ref{fig:5}.

The initial configuration is for a pulsed linac, focused on the needs of the neutrino program from the Main Injector, and uses the same 20 Hz repetition rate used for PIP-II injection into the Booster. It is also compatible with a CW linac. Acceleration to 2.4 GeV requires 1.4 GeV of acceleration from 1 GeV (or 1.6 GeV from 0.8~GeV). PIP-II cryomodules should be capable of at least 120 MeV. With additional R\&D focused on the 650 MHz cavities, we can expect improvements in both accelerating gradients and intrinsic quality factors. For this exercise we assume modest improvements of an accelerating gradient to 20.9 MV/m (which should give us an energy gain of 133 MeV per cryomodule) with a quality factor of $6\times10^{10}$. Thus, 11 cryomodules will be more than adequate to provide 1.4 GeV energy gain. Each cryomodule is 9.92 m long, and in PIP-II the cryomodules are in an 11.8 m lattice with a quadrupole doublet between each module.  For this linac a single quad per module would be used, so the lattice could be a bit more compact. 11 modules with the 11.8 m period requires 130 m, which fits into the 290 m slot, with 160 m of additional transport available, which can be used for optics matching and collimation, as well as elements to enable extraction into an external beam facility. The end of the linac could include pulsed kickers and/or RF separators to transport beam into a new intensity frontier facility in this vicinity, where pulsed or CW beam could feed a new family of experiments.

\subsection{Transport to final linac}

Following the 650 MHz linac the beamline must be bent and directed toward the RR-10 injection. The total bend is $105^\circ$, and a $\sim$165 m long transport is needed to ensure that magnetic stripping of 2.4 GeV H$^-$ does not occur. This would be obtained by a 8-cell FODO achromat ($720^\circ$ phase advance). The bend will include momentum collimation to reduce losses in the 1300 MHz linac and injection.

\subsection{1300 MHz pulsed linac}

After the $105^\circ$ bend, there is a 510 m long transport, which must include the 2.4 to 8 GeV 1300~MHz SRF linac plus optical matching and the final transport into the RR. The linac would contain LCLS-II style cryomodules, operating in pulsed mode. Each of these is 12.5 m long and includes 8 cavities and a focusing quadrupole magnet. Each cryomodule could provide at least 200 MeV of CW acceleration, with a mean accelerating gradient of 24 MV/m. 28 of these cryomodules would be needed, which implies a $\sim$350 m long linac. With some space reserved for matching optics, we can allot $\sim$410 m for the linac, leaving $\sim$100 m for the final transport matching into the RR. This is somewhat more space constrained than the 650 MHz linac.

However, if only pulsed mode of operation is considered, we can raise the accelerating gradient from 24 MV/m to 33.7 MV/m or even higher, and then obtain more than 280 MeV per cryomodule.  In that case, only 20 cryomodules will be required, which is reflected in Table~\ref{tab:iii} as our baseline scenario. This will significantly shorten the linac to $\sim$250 meters. We reserve an additional 50 meters in Table~\ref{tab:iii} for optical elements.

\subsection{Match into Recycler/MI}

We have reserved $\sim$150 m out of $\sim$200 m from the end of the linac for the transport into RR-10 injection. This section must include matching optics, including final bending into the RR, with the injection chicane magnets. Bending fields must be less than $\sim$0.05 T to avoid stripping. This 200 m section could also include RF separators or pulsed kickers into a beam line for a separate 8 GeV experimental facility. This new facility may prefer CW beam, which would require upgrading the 1300 MHz linac. Otherwise, a pulsed beam program using 20 Hz cycles not needed for the MI program could be developed.

\section{SRF cavities and cryomodules}

The demonstrated and projected performance of SRF cavities and systems has significantly changed since Project X. Two major discoveries at Fermilab have greatly improved SRF cavity performance \cite{5,6}. First, nitrogen doping of SRF cavities has been shown to reduce the BCS surface resistance below previously perceived limits. Second, effective magnetic flux expulsion by fast cooldown with a high thermal gradient has enabled achieving record low residual resistances. These two innovations, combined with continuing optimization of cavity treatments, have greatly increased usable gradients, with increased $Q$ values. Most recently, a 75/120$^\circ$C two-step low temperature bake of SRF cavities was shown to raise $Q$ by $\sim$50\%, and increased RF gradient to $\sim$50 MV/m for 1300 MHz cavities \cite{7}. These improvements will be incorporated into the BRL design and were considered in developing the SRF linac parameters presented in Table~\ref{tab:iv}. Continued SRF R\&D during the Booster Replacement program could result in even greater improvements.

\begin{table}[htbp]
\centering
\caption{\label{tab:iv} BRL SRF system parameters.}
\smallskip
\begin{tabular}{|l|l|l|}
\hline
\textbf{Parameter} & \textbf{650 MHz} & \textbf{1300 MHz}\\
\hline
Geometric $\beta$ & 0.92 & 1.0\\
Cells per cavity & 5 & 9\\
Beam aperture radius $a$ & 59 mm & 35 mm\\
$R/Q$ & 610 $\Omega$ & 1036 $\Omega$\\
$G=Q_0 \times R_s$ & 255 $\Omega$ & 270 $\Omega$\\
Accelerating gradient $E_{acc}$ & 20.9 MV/m & 33.7 MV/m\\
$E_{pk}$ & 43.4 MV/m & 67.5 MV/m\\
$B_{pk}$ & 81.5 mT & 140 mT\\
$Q_0$ & $6.0 \times 10^{10}$ & $2.0 \times 10^{10}$\\
$Q_L$ & $1.26 \times 10^{7}$ & $1.63 \times 10^{7}$\\
Beam current & 2 mA & 2 mA\\
Cavity fill time & 5.8 ms & 2.77 ms\\
Cavity dynamic losses at 2 K & 13.5 W & 59.1 W (pulsed)\\
Cavity RF power & 77.5 kW & 84 kW\\
Cavities per cryomodule & 6 & 8\\
Cryomodule length & 9.92 m & 12.5 m\\
Number of cryomodules & 11 & 20\\
\hline
\end{tabular}
\end{table}

\subsection{Cryomodule parameters}

\begin{figure}[htbp]
\centering 
\includegraphics[width=1.0\textwidth]{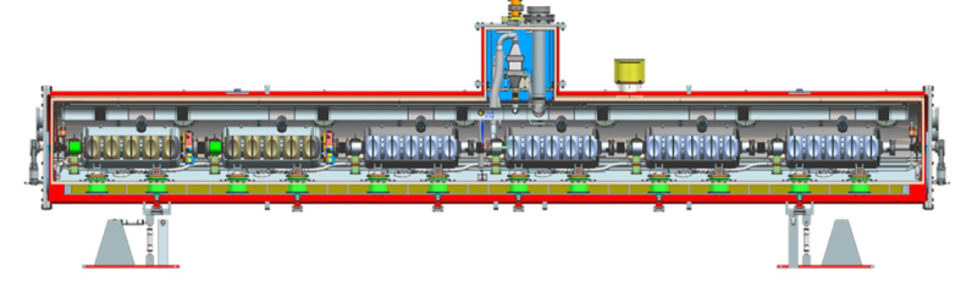}
\qquad
\caption{\label{fig:6} Cross section of a 9.92 m long 650 MHz cryomodule, containing six 5-cell RF cavities.}
\end{figure}

\begin{figure}[htbp]
\centering 
\includegraphics[width=1.0\textwidth]{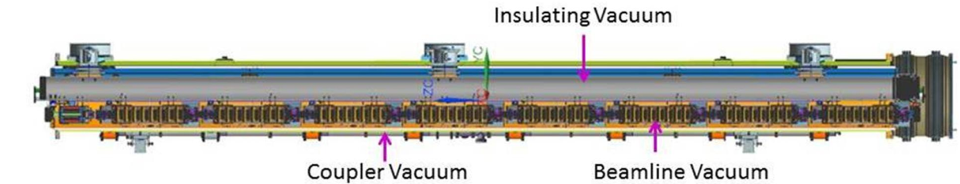}
\qquad
\caption{\label{fig:7} Cross section of a 12.5 m long 1300 MHz cryomodule, containing eight 9-cell cavities, a focusing magnet, following the LCLS-II design.}
\end{figure}

The building blocks for linac construction are the HB650 cryomodules, developed for PIP-II, and the 1300 MHz cryomodules, developed for the ILC (for pulsed operation) and the LCLS-II project at SLAC (for CW operation) \cite{8}. These designs are advanced and can be implemented for BRL with minimal modifications. Cross sections of the 650 MHz and 1300 MHz cryomodules are shown in Figs.~\ref{fig:6} and \ref{fig:7}, respectively.

The 650 MHz cryomodule contains six 1.061 m long cavities within a total length of 9.92~m. For PIP-II, an accelerating gradient of $E_{acc}$ = 18.8 MV/m yields an acceleration of 120 MeV per cryomodule. Upgrading this to 20.9 MV/m yields 133 MeV of acceleration, which should be available for the Booster Replacement linac with modest investments in R\&D. 

The 1300 MHz cryomodule contains eight 1.038 m cavities, which are included in a 12.5 m length. This would provide 200 MeV at 24 MV/m (slightly improved LCLS-II-HE technology). An upgrade to 33.7 MV/m (R\&D on ILC technology, currently in progress) would increase that to 280 MeV, which is what is used in our initial scenario. Future study would determine an optimum gradient. The gradient must not be above thresholds for H$^-$ stripping.

\subsection{SRF cavities}

\begin{figure}[htbp]
\centering 
\includegraphics[width=1.0\textwidth]{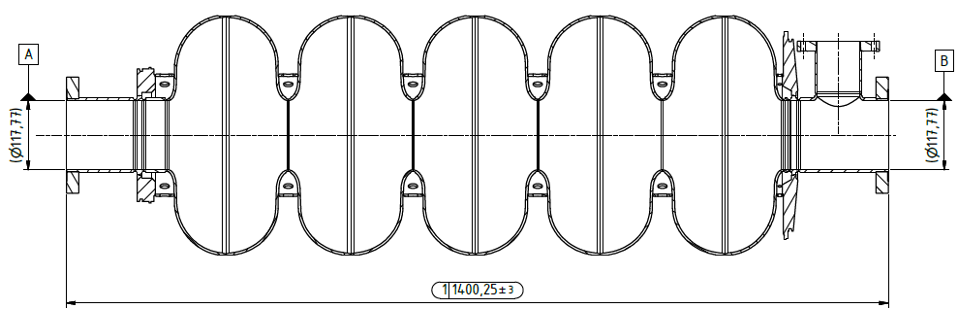}
\qquad
\caption{\label{fig:8} Drawing of the HB650 cavity.}
\end{figure}

The PIP-II HB650 cavity ($\beta_G$ = 0.92) is a 5-cell elliptical SRF cavity operating at 650 MHz. For the BRL 1 to 2.4 GeV linac we assume that the cavity will provide an energy gain of 22.2 MeV to H$^-$ beam. The cavities will operate with an intrinsic quality factor of $6.0\times10^{10}$ at 2 K. The cell shape optimizes the ratios of the peak surface magnetic and electric fields, $H_{pk}$ and $E_{pk}$, to the accelerating gradient, $E_{acc}$. The cavity iris aperture was chosen at 118 mm as a compromise between the peak field ratios, shunt impedance, cell-to-cell coupling, and cavity handling during surface processing. The beam tube has penetrations for the fundamental power coupler (FPC), which is centered at 96~mm from the end iris and for the pickup probe at the other end of the structure. An external quality factor of the FPC is $Q_{ext} = 1.2\times10^7$.  Fig.~\ref{fig:8} show a mechanical drawing of the HB650 cavity. See Table~\ref{tab:iv} for a listing of the HB650 cavity electromagnetic parameters. If necessary, a $\beta_G$ = 1 version of the HB650 cavity can easily be derived from the existing design.

\begin{figure}[htbp]
\centering 
\includegraphics[width=1.0\textwidth]{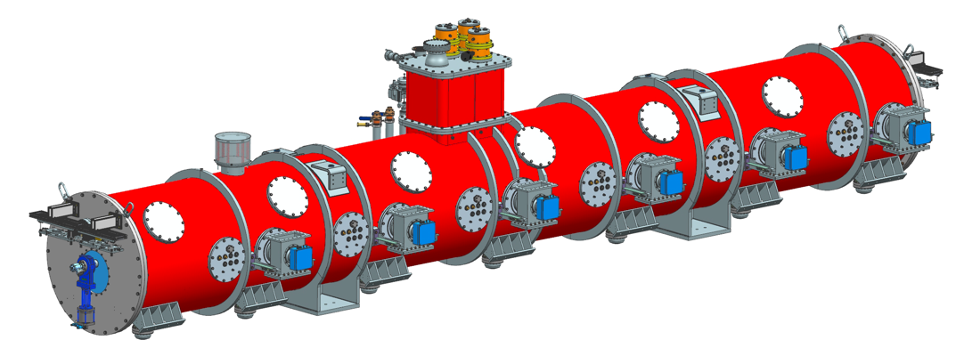}
\qquad
\caption{\label{fig:9} HB650 cryomodule \cite{28}.}
\end{figure}

The cavity is built from 4 mm thick high RRR niobium sheet material via stamping and electron beam welding (EBW). Vacuum seals on all demountable interfaces are achieved via NbTi flanges using AlMg diamond seal gaskets. The cavity is stiffened with 100 mm inner diameter high RRR 4 mm thick niobium stiffening rings. NbTi transition spools, which are stiffened by the same rings, provide the interface to the helium vessel and cavity tuner. The transition spools are EB welded to the beam tubes on each end group. Grade 2 titanium end cans, bellows, and helium vessel, which are TIG welded to the transition rings and to each other, form the liquid helium enclosure. A cylindrical helium vessel will be furnished with two helium inlets for fast cavity cooling, with a chimney to remove up to 33 W of average dissipated power as well as positioning and tuner lugs. All auxiliary components and the cryomodule are identical to the PIP-II HB650 cryomodule shown in Figs.~\ref{fig:6} and \ref{fig:9}. The HB650 cryomodule can operate in either CW or pulsed mode.

The BRL 2.4 to 8 GeV linac design is based on the LCLS-II/LCLS-II-HE cryomodule design (Fig.~\ref{fig:7}) \cite{8}. As, in turn, the latter design is based on the European XFEL cryomodules, it is capable to operate in either CW or pulsed mode. The 9-cell TESLA shape cavities \cite{9} are a de~facto standard for SRF accelerators operating at 1300 MHz. Here we assume that with further R\&D progress, the 9-cell TESLA SRF cavities \cite{9} will be able to operate at 33.7 MV/m at $Q_0 = 2\times10^{10}$ (compared to the ILC baseline of 31.5 MV/m at $1\times10^{10}$). Some cavity parameters are listed in Table~\ref{tab:iv}.

\section{H$^-$ linac constraints: Beam losses}

Beam losses can be a significant limitation in a high intensity linac. To keep the radioactivation of the beam line components low enough for ``hands-on maintenance'', activation levels must be below $\sim$100 mrem/hr at 30 cm from a component surface, after extended operation. From previous experience this implies losses of less than $\sim$1 W/m. A safety limit of $\sim$0.2 W/m would allow relatively unrestricted maintenance \cite{10,11}.

\subsection{Magnetic stripping constraints}

The 8 GeV linac beam must be transmitted as H$^-$, for compatibility with H$^-$ injection into the Recycler or Main Injector, and the bending fields in the BRL transport lines must be limited to avoid magnetic stripping to H$^0$ \cite{11}. The 8 GeV linac has three locations with significant amounts of bending magnets: the initial bend of $\sim45^\circ$ following the PIP-II linac where the beam has an energy of ~1 GeV, the bend of $\sim105^\circ$ at the end of the 3 GeV CW linac, and smaller bends at 8 GeV associated with injection into the Recycler/Main Injector.

The stripping length can be estimated using this formula of Schrek \cite{12,13}:

\begin{equation}
L_{\mbox{strip}} = \beta \gamma c \tau = \beta \gamma c \frac{a}{3.197 B_t p} \exp \left( \frac{b}{3.197 B_t p} \right) \mbox{ meters,}
\end{equation}
where $p$ is the H$^-$ momentum, $B_t$ is the magnetic field and $a$ and $b$ are parameters fitted from data. Keating et al. \cite{14} obtained $a = 3.073\times 10^{-14}$ and $b = 44.14$ from 800 MeV data.

\begin{figure}[htbp]
\centering 
\includegraphics[width=.7\textwidth]{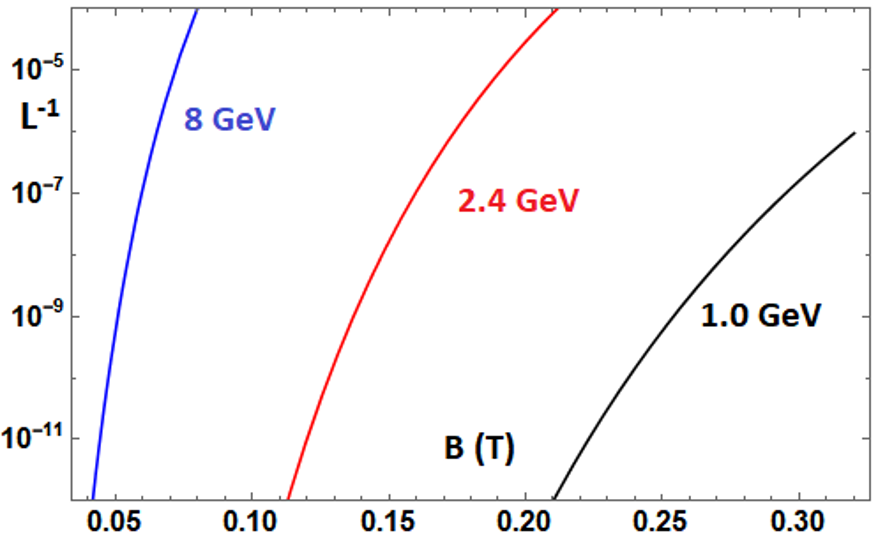}
\qquad
\caption{\label{fig:10} Magnetic stripping rate (measured in m$^{-1}$) as a function of $B$ (in T) for 1, 2.4 and 8 GeV H$^-$. }
\end{figure}

For 1 GeV protons the transport is a mirror image of the PIP-II transport to the Boost-er. For that transport, the PIP-II design set a limit of 0.277 T, at which $\tau$ = 0.12 s, and $L = 6.43\times10^7$~m. The BRL transport would have the same values. Losses per meter would be $1.6\times10^{-8}$, which corresponds to 0.032 W/m at 2 MW beam power. The $60^\circ$ arc requires $\sim21.4$2~m of bend, which must be included in an achromatic lattice. The total losses would be $\sim3.5\times10^{-7}$, which is relatively small. Fig.~\ref{fig:10} provides plots of the magnetic stripping rates as a function of magnetic field for the three energies of interest. Requiring $L^{-1} \sim 10^{-8}$~m$^{-1}$ yields $B \cong$ 0.27, 0.15, and 0.055 T for $E_{H^-} =$~1, 2.4, and 8 GeV, respectively.

\subsection{Electron stripping}

A related question is whether the H$^-$ ions could be stripped by the acceleration cavity fields, and whether that sets a limit on the cavity maximum field that is lower than other gradient limits. In a periodic acceleration structure having large aperture and a phase velocity $\beta$ = 1, the axial electric field distribution on the axis is close to sinusoidal. This means that the maximal longitudinal electric field $E_{z,max}$ on axis is two times higher than the acceleration gradient $E_{acc}$:

\begin{equation}
E_{z,max} = 2E_{acc}.
\end{equation}

According to Maxwell equations, in the paraxial area the RF magnetic field has only an azimuthal component, which is equal to:

\begin{equation}
B_{\varphi}(r,z) = -\frac{i\omega}{c^2}\frac{r}{2}E_{z}(z),
\end{equation}
where $\omega$ is the RF angular frequency. The maximal magnetic field amplitude is

\begin{equation}
B_{\varphi, max}(r,z) = \frac{\omega}{c^2}\frac{r}{2}E_{z,max}=\frac{r\omega}{c^2}E_{acc}.
\end{equation}

There is no stripping of an 8 GeV beam caused by the RF magnetic field if

\begin{equation}
B_{\varphi, max}(r,z) = \frac{r\omega}{c^2}E_{acc} \leq B_0 = 0.055 \mbox{ T}.
\end{equation}

\begin{figure}[htbp]
\centering 
\includegraphics[width=.7\textwidth]{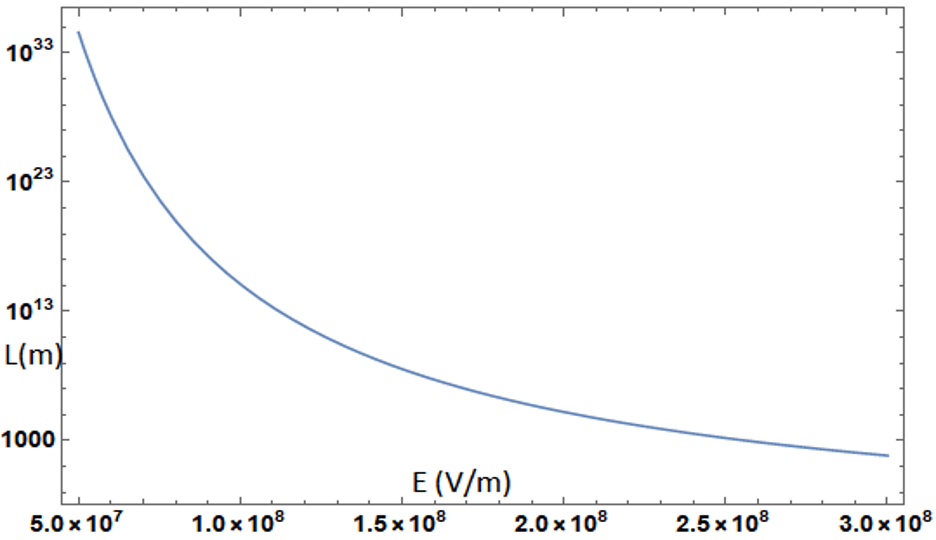}
\qquad
\caption{\label{fig:11} Stripping length as a function of accelerating gradient for 8 GeV H$^-$. At 50 MV/m the mean stripping length is greater than $10^{34}$ m. This is reduced to $3.5\times10^8$ m at E = 150 MV/m.}
\end{figure}

\noindent
Therefore, the acceleration gradient limitation is

\begin{equation}
E_{acc} < \frac{c^2}{r\omega}B_0 = \frac{\lambda}{2\pi r}cB_0,
\end{equation}
where $\lambda$ is the RF wavelength.
The stripping length for H$^-$ as a function of accelerating gradient is shown in Fig.~\ref{fig:11}.

In the beam frame the magnetic field transforms into an electric field with a magnitude of $\beta \gamma cB_\varphi$. The rms beam emittance is $\sim0.3$ mm-mrad (normalized), which places the 8 GeV rms beam size at $\sim1$ mm, at a large $\beta_\bot$ = 30 m. If the beam particles have amplitudes less than $\sim1$~cm, the 0.055 T limit on the transverse magnetic field used in the bending magnets transforms to a limit of $\sim60$ MV/m. Provided that particle amplitudes can be confined within < 1 cm, magnetic and electric field stripping in the SRF cavities should not be a significant problem, particularly if $E_{acc}$ is less than 40 MV/m.

\subsection{Intra-beam scattering}

An unexpected beam loss mechanism in the H$^-$ linac at SNS was identified by V. Lebedev as due to neutralization from intra-beam stripping. The Booster Replacement Linac is also H$^-$ and is therefore vulnerable to this loss mechanism. An equation for the beam loss due to intra-beam stripping is presented in \cite{15}:

\begin{equation}
\frac{1}{N}\frac{dN}{dt} = \frac{N\sigma_{max}\sqrt{\sigma_{vx}^2+\sigma_{vy}^2+\sigma_{vz}^2}}{8\pi^2 \sigma_x \sigma_y \sigma_z}F\left(\sigma_{vx},\sigma_{vy},\sigma_{vz} \right),
\end{equation}
where $\sigma_{max} \simeq4\times10^{-15}$ cm$^{-2}$, $N$ is the number of H$^-$ ions/bunch, $\sigma_x$, $\sigma_y$, $\sigma_z$ are beam sizes in the beam frame, $\sigma_{vx}$, $\sigma_{vy}$, $\sigma_{vz}$ are the beam velocity spreads, and

\begin{equation}
F\left(a,b,c \right) = \frac{1}{\pi}\sqrt{\frac{x^2}{a^2}+\frac{y^2}{b^2}+\frac{z^2}{c^2}}\exp\left( -\frac{x^2}{a^2}-\frac{y^2}{b^2}-\frac{z^2}{c^2} \right) \frac{dxdydz}{abc}.
\end{equation}

$N$ is $\sim1.9\times10^8$ at 5 mA peak current, $\sigma_x = \left( \epsilon_{n,x}  \beta_x / \beta_\gamma \right)^{1/2}$, $\theta_{x,rms} = \left( \epsilon_{n,x}/\frac{\beta_x}{\beta_\gamma} \right)^{1/2}$, $\theta_{\parallel,rms} = \delta p/p$, and $\sigma_s$ is the bunch length. The integrated function $F$ is close to 1 at Project X parameters. At typical parameters $\epsilon_n = 0.3\times10^{-6}$ m-rad, $\beta_{x,y}$ = 10 m, $\gamma$ = 2.07 to 9.53 (1 to 8~GeV), $\sigma_s$ = 1.5 mm, $\theta_\parallel$ = $\sim0.0003$, we get values of  
$\frac{1}{N}\frac{dN}{ds}$
from $\sim 4\times10^{-8}$ at 1 GeV to $\sim 2\times10^{-8}$ at 8 GeV. This corresponds to 0.04 to 0.02 W/m at 1 MW. A more complete evaluation of these was made by Ostiguy for Project X, with evaluation of beam sizes from tracking through the lattice~\cite{16}. Those results were similar to the present evaluation.

\subsection{Black-body radiation stripping}

The beam pipe would be filled with low-energy photons from thermal black-body radiation. At room temperature (300 K), $kT$ = 0.02587 eV and the spectral energy distribution peaks at $\sim0.06$~eV. A much larger exchange of $E_0$ = 0.754 eV is needed to ionize H$^-$ at rest. The photons are Doppler shifted by a factor of up to 2$\gamma$  in the H$^-$ ion rest frame at high energies. At 8 GeV, the peak is shifted above that threshold and H$^-$ stripping can occur \cite{10}.

The photo-detachment cross section is

\begin{equation}
\sigma(E^\prime) = 8\sigma_{max} \frac{E^{3/2}\left( E^\prime - E_0 \right)^{3/2}}{E^{\prime3}},
\end{equation}
where $\sigma_{max} = 4.2\times10^{-21}$ and $E^\prime$ is the photon energy in the H$^-$ rest frame:

\begin{equation}
E^\prime = \gamma (1 + \beta \cos \alpha).
\end{equation}

The stripping rate can be calculated using the following equation:

\begin{equation}
\frac{1}{L} = \frac{8\sigma_{max}E_0^{3/2}}{2\pi^2 \beta \gamma^3(\hbar c)^3}\int_{E_0}^{\infty}dE^\prime \int_{-1}^{+1}du\frac{1}{(1+\beta u)^2}\frac{(E^\prime - E_0)^{3/2}}{E^\prime}\frac{1}{[\exp{(E^\prime / kT\gamma(1+\beta u))}-1]}.
\end{equation}

We evaluated this expression to be $\sim7.8\times10^{-7}$ m$^{-1}$, in good agreement with Carneiro et~al.~\cite{10}. This is a fairly large value. The pulsed version of the 8 GeV beam ($\sim$200 kW) would have 0.16~W/m while a CW version at 2 MW would have 1.56 W/m. The transport at 8 GeV is relatively short, so the resulting beam loss should be manageable ($1.56\times10^{-5}$ in 20 m). The radiation stripping can be greatly reduced by cooling the beam pipe to a lower temperature, which reduces the photon energy spectrum proportionately. A reduction to 150 K (from liquid nitrogen cooling) would reduce losses to $\sim2.5\times10^{-8}$, enabling easier maintenance and more manageable CW operation.

\subsection{Beam-gas stripping}

Collisions of H$^-$ with background gas molecules can strip the H$^-$ ions, causing beam loss \cite{10}. The lifetime $\tau_m$ of an H$^-$ ion in the presence of residual gas is given by

\begin{equation}
\tau_m = \frac{1}{\beta c d_m \sigma_m},
\end{equation}
where $d_m$ is the gas particle density and $\sigma_m$ is the interaction cross section. The beam fraction loss per unit length is

\begin{equation}
\frac{1}{L} = \frac{1}{\tau_m \beta c},
\end{equation}

This is to be summed over gas components. If we assume the gas is``air'', then $\sigma_m = \sim0.65\times10^{-18}$ cm$^2$ and $d_m = 3.2\times10^{22} P$ (Torr) m at $T$ = 300 K then $L^{-1} = 2.1 P$ (Torr) m$^{-1}$. With a vacuum of $10^{-8}$ Torr, losses are $2.1\times10^{-8}$ m$^{-1}$, or 0.042 W/m for a 2 MW beam. \cite{12} used a generic vacuum of 70\% H$_2$, 10\% H$_2$O, 10\% CO$_2$, 10\% CO), which would make the average value of $\sigma_m \simeq0.15\times10^{-18}$ and reduce the losses by a factor of $\sim4$ at the listed pressure.

\section{MI/RR injection}

The locations of PIP-II and the Main Injector, as well as the requirement that MI extraction feed the LBNF beamline, restrict the possible locations for H$^-$ injection. Charge exchange injection requires modification of a long straight section. MI-10 is the only long straight section that readily accommodates a linac from PIP-II to the MI. The next possibility, MI-30, is another km downstream, and requires extra bending which would strip H$^-$ ions. The previous straight section, MI-60, would either inject into the MI in the wrong direction or require a $\sim180^\circ$ bend that would strip H$^-$ ions. However, MI-10 has been chosen as the straight section for extraction to LBNF, and the extraction kickers and septum magnets will occupy the space needed for injection to the MI \cite{18}. Unless MI extraction is changed, MI-10 cannot be used for charge exchange injection.

\begin{figure}[htbp]
\centering 
\includegraphics[width=1.0\textwidth]{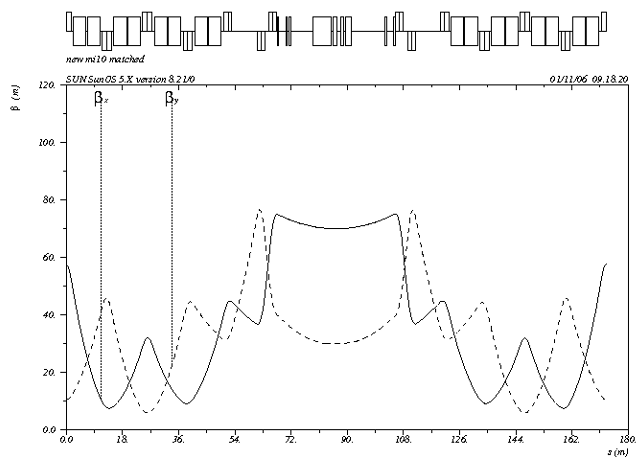}
\qquad
\caption{\label{fig:12} Lattice with betatron functions for the RR-10 straight section, adapted for foil injection. A $\sim$36 m segment between focusing quads is reserved for injection kickers, injection bump magnets and stripping foil.}
\end{figure}

The Recycler Ring lies directly above the MI, and its corresponding straight section, RR-10 can be modified to accommodate charge exchange injection, as shown in Fig.~\ref{fig:12}. The Recycler Ring can then be used to accumulate multiple pulses from the linac to be followed by single turn transfer to the MI at MI-22 or MI-30. That general injection procedure was included in the Project~X plan and was developed in some detail by D. Johnson et al. \cite{11} and A. Drozhdin et al. \cite{19}. That injection is the baseline injection plan considered in this article, modified to match the present MI conditions.

\subsection{Foil injection}

\begin{figure}[htbp]
\centering 
\includegraphics[width=1.0\textwidth]{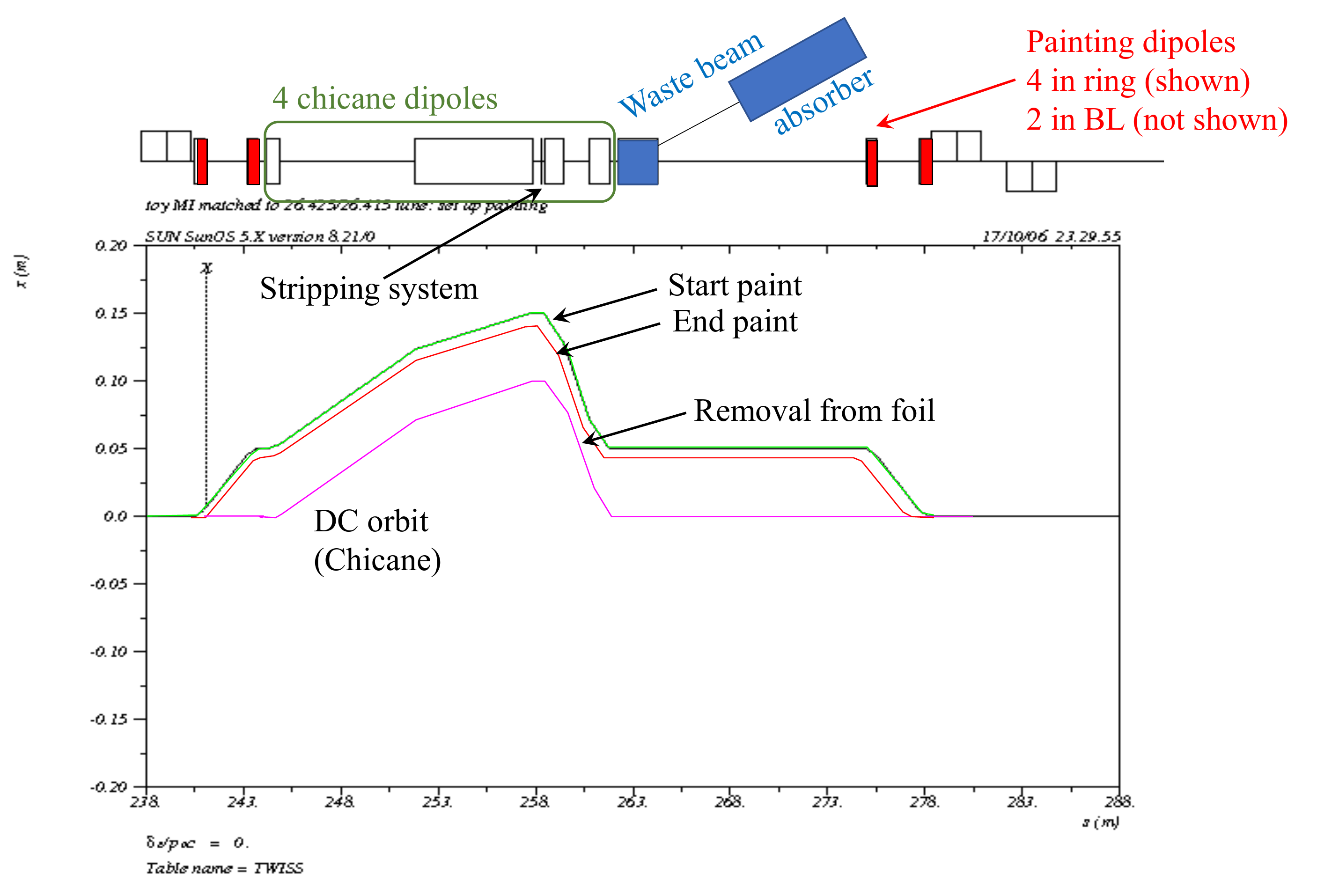}
\qquad
\caption{\label{fig:13} Injection insert components with kickers, and orbit variations used for injection with painting on the foil.}
\end{figure}

The baseline injection method from the linac into the RR is multiturn foil-stripping injection. During injection the beam orbit is displaced toward the foil by a 4-dipole chicane (see Fig.~\ref{fig:13}). The long second dipole (with a low field of $B$ = 0.055 T) is shared by both circulating beam (H$^+$ or protons) and the injected H$^-$ beam to converge on the foil between the middle dipoles. The smaller H$^-$ beam strikes the foil in a corner of the circulating proton beam distribution, where the H$^-$ ions are stripped to H$^+$ to join the circulating beam distribution. The painting dipole fields are varied such that the injected H$^+$ populate the circulating distribution matched to the enlarged final beam emittance, while minimizing the number of foil hits per proton.

\begin{figure}[htbp]
\centering 
\includegraphics[width=0.8\textwidth]{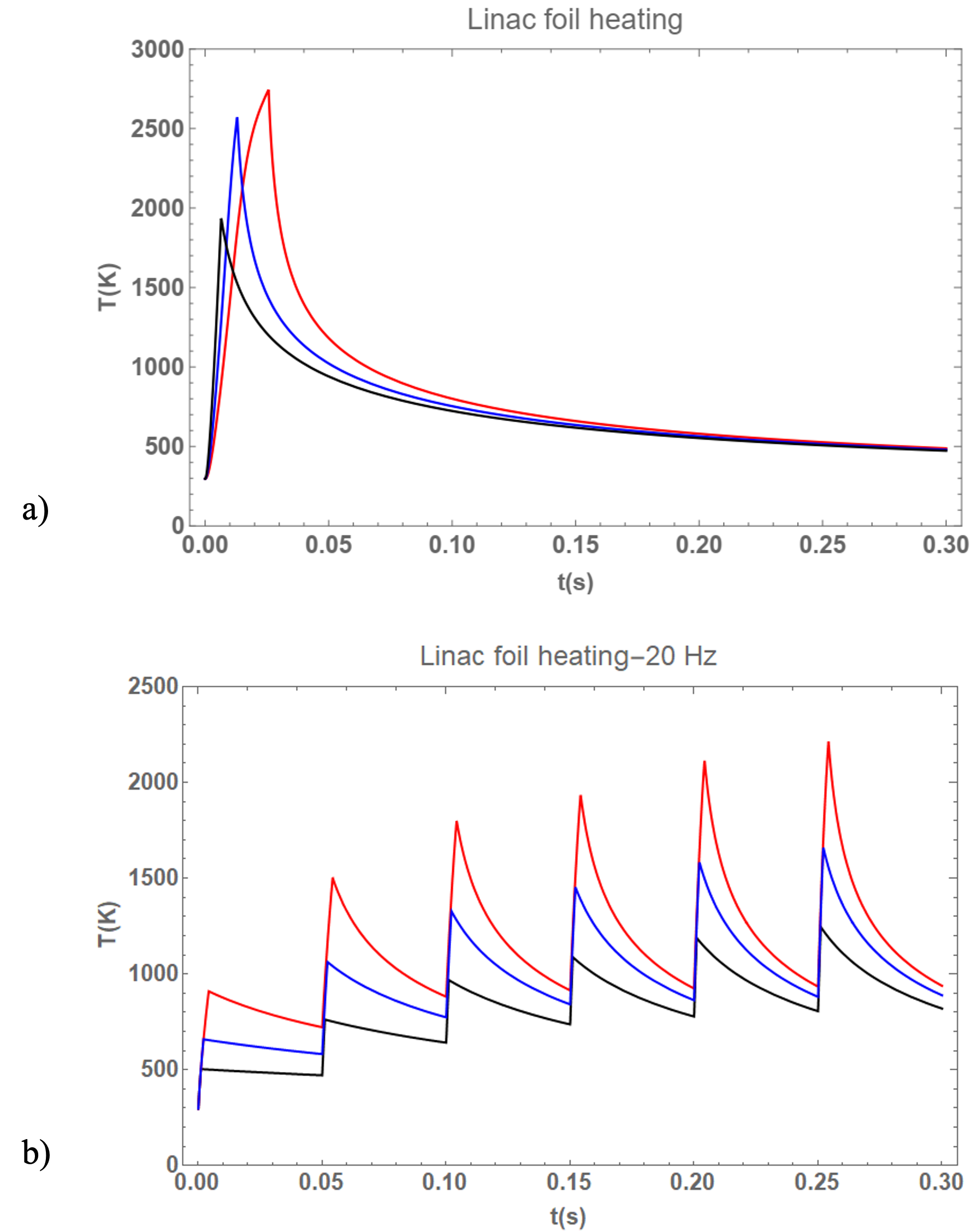}
\qquad
\caption{\label{fig:14} a) Foil heating with single 26 mA-ms injection at 1 mA (red), 2 mA (blue) and 4 mA (black). b) Foil heating in 6-step injection (4.4 mA-ms/step) at 1 (red), 2 (blue) and 4 mA (black).}
\end{figure}

High intensity multiturn H$^-$ injection into the RR/MI, with injection painting and foil heating, was simulated by Drozhdin et al. \cite{19} and further explored by Neuffer \cite{20}. Injection requires $\sim26$~mA-ms of beam. At 1 -- 2 mA, this implies 2300 -- 1150 turns. If this were injected in a single pulse, the foil would heat to $\sim2500$ K, which is unacceptably high. (Unacceptable sublimation would occur at $\sim1800$ K.) The preferred injection procedure is to split the injection into a number of separate shorter injections, spaced by the pulsed linac rep rate, and then sequentially inject into the ring, while following a foil painting program to minimize the number of foil hits. Fig.~\ref{fig:14} displays calculations of foil heating in a 6-step injection at 1, 2, 4 mA currents, which reduces peak $T$ to 2200, 1660, and 1250 K, respectively.  The 2 and 4 mA numbers are acceptable. Parameters of injection scenarios are presented in Table~\ref{tab:v}.

\begin{sidewaystable}[htbp]
\centering
\caption{\label{tab:v} Summary of injection scenario parameters. In the first three cases the RR is filled with a single long pulse of 1 mA, 2 mA, and 4 mA, respectively. In the last three injection is split into 6 separate (20 Hz) pulses.}
\smallskip
\begin{tabular}{|l|l|l|l|l|l|l|}
\hline
\textbf{Scenario} & \multicolumn{3}{c|}{\textbf{Single long pulse}} & \multicolumn{3}{c|}{\textbf{6 separate pulses}}\\ 
\cline{2-7}
 & 1 mA - 26 ms & 2 mA & 4 mA & 1 mA - 4.3 ms & 2 mA & 4 mA\\ 
\hline
Injection time & 25.72 ms & 14.8 ms & 6.9 ms & 6$\times$4.28 ms & 6$\times$2.2 ms & 6$\times$1.1 ms\\
Number of turns & 2334 & 1167 & 584 & 6$\times$395 & 6$\times$198 & 6$\times$99\\
Ave. hits per proton & 120 & 61 & 33 & 120 & 60 & 32\\
Max. hit density & $5.15\times10^{14}$ & $2.60\times10^{14}$ & $1.33\times10^{14}$ & $5.14\times10^{14}$ & $2.58\times10^{14}$ & $1.31\times10^{14}$\\
Maximum foil $T$ & 2750 K & 2580 K & 1930 K & 2215 K & 1660 K & 1250 K\\
\hline
\end{tabular}
\end{sidewaystable}

\subsection{Beam losses at injection}

\begin{figure}[htbp]
\centering 
\includegraphics[width=.8\textwidth]{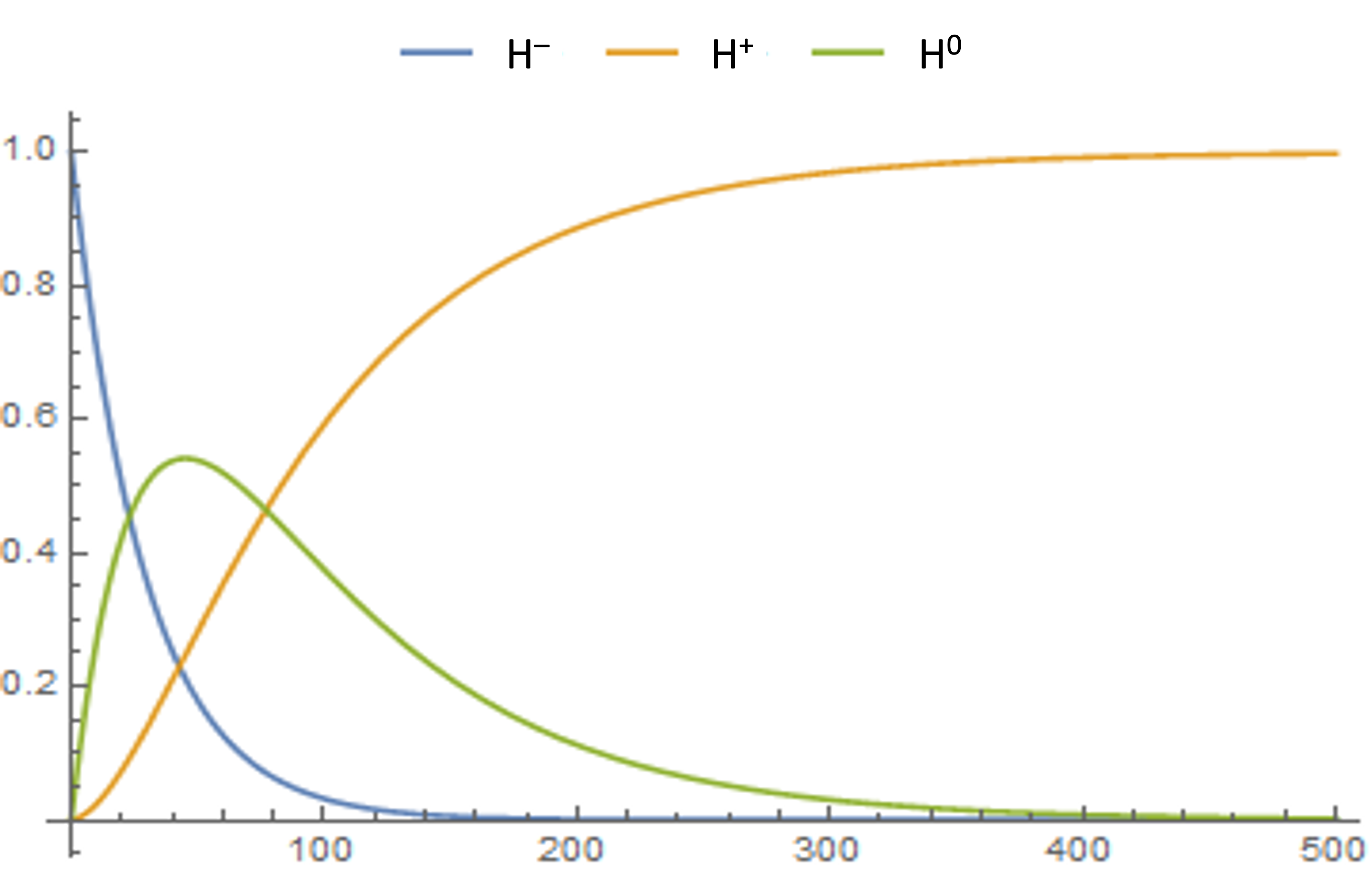}
\qquad
\caption{\label{fig:15} Fraction of beam that is H$^-$, H$^0$, or H$^+$ as it passes through a C foil with final thickness of 500~$\mu$g/cm$^2$. At 400 $\mu$g/cm$^2$, the beam is ~96.4\% H$^+$, and 3.6\% H$^0$. At 500 $\mu$g/cm$^2$, it is ~98.6\% H$^+$. At 600~$\mu$g/cm$^2$, it would be ~99.5\% H$^+$.}
\end{figure}

In H$^-$ injection the beam passes through a foil, where interactions with the foil material causes the ion to lose electrons, eventually being stripped to H$^+$ (protons) that can be stored and accelerated. The H$^-$ ions are stripped to H$^0$ and H$^+$, and H$^0$s are stripped to H$^+$. Equations for stripping versus foil thickness have been developed by Gulley et al. \cite{21}, from fits to measured stripping data. The equations are \cite{22}

\begin{equation}
f_{H^-}(t,\beta) = \exp[-(0.479+0.0085)\cdot 0.05t/\beta^2],
\end{equation}
\begin{equation}
f_{H^0}(t,\beta) = \frac{0.479}{0.479+0.0085-0.187} \left\{ \exp[-(0.187)\cdot 0.05t/\beta^2] -\exp[-(0.479)\cdot 0.05t/\beta^2] \right\},
\end{equation}
\begin{equation}
f_{H^+}(t,\beta) = 1 - f_{H^0}(t,\beta) - f_{H^-}(t,\beta),
\end{equation}
where $\beta = v/c$ is the usual kinematic factor for the incident H$^-$, $t$ is the carbon foil thickness in $\mu$g/cm$^2$.  For a 500 $\mu$g/cm$^2$ thick foil, 98.6\% of initial H$^-$ are stripped to H$^+$ (protons). For graphite (at density $\rho$ = 2.0 g/cm$^3$), this is a 2.5 $\mu$m thick foil, or 1.4 $\mu$m thick for diamond ($\rho$ = 3.6 g/cm$^3$). Fig.~\ref{fig:15} shows the variation of ion fraction through a foil with thickness of 500 $\mu$g/cm$^2$, which will be our reference design thickness.

Unstripped H$^-$ and H$^0$, including injected beam that misses the foil, are magnetically separated from the circulating beam in the dipole just downstream of the foil, stripped by a thicker downstream foil and directed into a beam absorber, designed to localize losses in a shielded enclosure. This enclosure and absorber will require new construction downstream of RR-10, and it will be constrained by the proximity of the MI extraction beamline and the MI and RR rings. From previous designs, we expect $\sim$2\% of the incoming beam to be lost, with ~90\% of that captured into the absorber; $\sim$0.2\% will be lost in a relatively uncontrolled manner in the near vicinity of extraction or further downstream in the RR. Future detailed design effort must determine that these losses are adequately managed. The injected beam and the accumulating proton beam will pass through the foil, losing energy and scattering, with energy straggling. This can lead to beam losses from rms emittance growth and large angle scattering events. 

Table~\ref{tab:vi} summarizes the various loss mechanisms, estimates of expected losses, and mitigation strategies for the BRL, including linac and injection losses \cite{17}. The large angle scattering and nuclear losses are based on Project X calculations that assumed each event caused beam loss and are expected to be an overestimate, and also do not separate out losses controlled by collimation and the injection beam dump. They should be recalculated in a more complete format.

\begin{table}[htbp]
\centering
\caption{\label{tab:vi} Loss mechanisms, expected effects and mitigation.}
\smallskip
\begin{tabular}{|l|l|l|l|l|}
\hline
\textbf{Loss process} & \textbf{Key parameters} & \textbf{Loss} & \textbf{Estimated} & \textbf{Mitigation}\\
   &  & \textbf{per meter} & \textbf{losses} & \textbf{strategies}\\
\hline
Magnetic stripping & $B$ (1 GeV) < 0.28 T & $1.6\times 10^{-8}$ & $3\times10^{-7}$ & Limiting $B$ field\\
 & $B$ (2.4 GeV) < 0.15 T & $2\times 10^{-8}$ & $3\times10^{-6}$ & \\
 & $B$ (8 GeV) < 0.055 T & $10^{-8}$ & $10^{-7}$ & \\
Black-body radiation & 1 GeV & $3.7\times 10^{-8}$ & $10^{-6}$ & Shorter, shielded \\
 & 2.4 GeV & $1.0\times 10^{-7}$ & $1.6\times10^{-5}$ & transport\\
 & 8 GeV, $T$ = 300 K & $7.8\times 10^{-7}$ & $\sim10^{-6}$ & Cooled beam pipe\\
Beam-gas interactions & $P\sim10^{-8}$ Torr & $2.1\times 10^{-8}$ & $<\sim 10^{-5}$ & Vacuum\\
Intra-beam stripping & $N = 2\times10^{8}$/bunch & $2-4\times10^{-8}$ & $<\sim 10^{-5}$ & Short transport\\
Foil: beam misses & 500 $\mu$g/cm$^2$ C foil &  & $\sim$2\% & Collimation before\\
 &  &  &  & foil matching,\\
 &  &  &  & absorber\\
Foil: H$^0$ &  &  & $\sim$1\% & Injection absorber\\
Foil: large-angle & 40$\pi$ mm-mrad &  & < 300 W & Collimation, \\
scattering & acceptance &  &  & reduce foil \\
 &  &  &  & crossing \\
Foil: nuclear & $L_N$ = 60 - 86 g/cm$^2$ &  & < 60 W & Collimation,\\
interaction &  &  &  & shielding\\
\hline
\end{tabular}
\end{table}

\section{Variant injection scenario}

The Recycler-based RR-10 injection provides the most direct injection from the BRL into the MI tunnel, but is limited by the restricted geometry of the MI-10 region, and requires accumulation in the Recycler Ring and transfer to the MI. The use of the Recycler also fixes the injection energy at 8 GeV. A direct injection into the MI would enable full use of the larger MI aperture and could accommodate different injection energies. However, the direct injection into MI-10 is not possible without changing the current design for the LBNF, which uses MI-10 for extraction \cite{18}. Other MI straight sections are not readily available for injection from the BRL, as described in section~7.

\begin{figure}[htbp]
\centering 
\includegraphics[width=.9\textwidth]{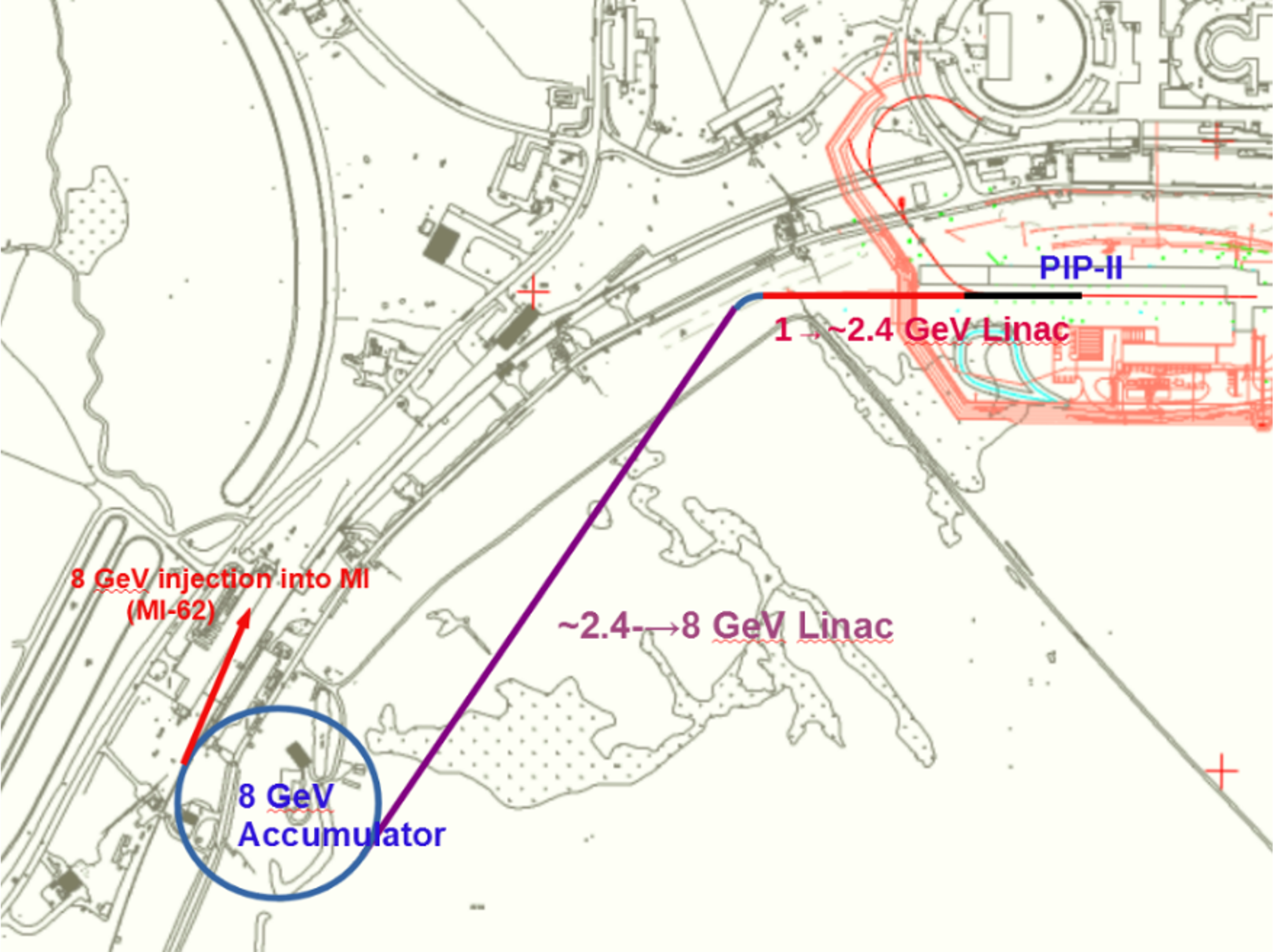}
\qquad
\caption{\label{fig:16} Alternative layout of BRL. An 8 GeV linac leads to an accumulator ($\sim1/6$ MI circumference), where foil stripping of H$^-$ beam over $\sim4.2$ ms-mA pulses can occur. }
\end{figure}

An alternative injection scenario is displayed in Fig.~\ref{fig:16}. The linac is reconfigured to inject into a new 8 GeV storage ring, which is placed inside the Main Ring near MI-60. The new ring would be designed with an acceptance larger than the MI, would have an injection straight optimized for multiturn foil injection from the linac and an extraction matched into the Recycler or MI at MI-60 or MI-62. The ring is $\sim1/6$ of the MI in circumference; 6 box-car stacked pulses from the ring would fill the RR or MI in each MI cycle \cite{24}. 

This scenario requires the construction of a new storage ring, which can have an injection system fully optimized to limit and control losses. In cycles not needed for MI-DUNE, the storage ring could also accumulate 8 GeV beam for other experiments, including neutrino beams, pulsed pion and muon beams. The stored beam may also be needed for the Fermilab high-intensity program that will complement the DUNE-based high-energy neutrino program.

\section{BRL power requirements}

The AC power requirements have been estimated for the pulsed operation of the two linacs. The AC power demand has two main contributions: from the cryogenic heat loads of the SRF linacs (due to power dissipation in SRF cavities, fundamental power coupler heat loads, and static heat loads) and from the RF power system. The cryogenic heat loads were calculated for the three temperatures at which heat will be extracted either with liquid helium or with appropriate thermal shields and heat intercepts. These temperatures are 2 K, and in the ranges of 5 -- 8 K, and 40 -- 80 K. Uncertainty factors of 30\% and 10\% were added to the calculated static and dynamic heat loads, respectively, to provide safety margins. The following inverse coefficients of refrigeration performance were used to calculate the required AC power: 790 W/W for 2 K, 208 W/W for 5 -- 8 K, and 21.4 W/W for 40 -- 80 K. For the RF power, we added a 10\% allocation for losses in the RF distribution system and 6\% for overhead. A 50\% efficiency was used to convert RF power to AC power. Table~\ref{tab:vii} summarizes the main contributions for the two SRF linacs.
From this, one can calculate an AC-to-beam power efficiency of the linacs. The 650 MHz linac delivers 0.14 MW to the beam, while the 1300 MHz linac delivers 0.49 MW. Thus, the overall power efficiency of the two linacs is 0.63/4.63 = 13.7\%.

\begin{table}[htbp]
\centering
\caption{\label{tab:vii} BRL AC power.}
\smallskip
\begin{tabular}{|l|r|r|}
\hline
\textbf{Parameter} & \textbf{650 MHz} & \textbf{1300 MHz}\\
\hline
2 K circuit power$^1$ & 180 kW & 526 kW\\
5 -- 8 K circuit power$^1$ & 88 kW & 117 kW\\
40 -- 80 K circuit power$^1$ & 70 kW & 129 kW\\
RF system power$^2$ & 940 kW & 2.58 MW\\
\hline
\textbf{Grand total} & \textbf{1.28 MW} & \textbf{3.36 MW}\\
\hline
\end{tabular}
\end{table}
 
{\small $^1$with 30\% uncertainty for static heat loads and 10\% uncertainty for dynamic heat loads.}

{\small $^2$with 10\% allocation for losses in the RF system and 6\% overhead.}

\section{Proton Intensity Upgrade scenario}

\begin{figure}[htp]
\begin{centering}
\includegraphics[width=\textwidth]{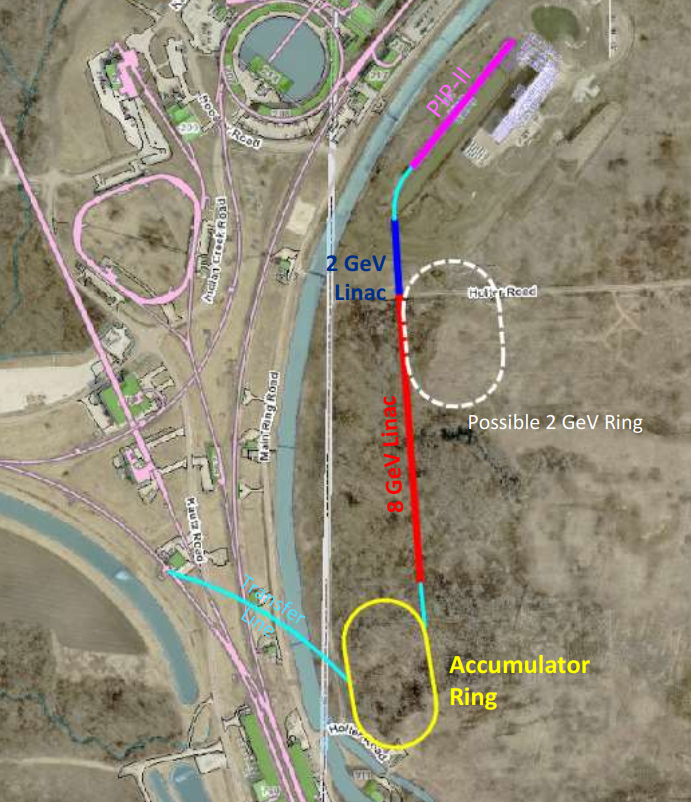}
 \caption{Schematic overview of a possible layout for the 8 GeV linac PIU scenario. Following the PIP-II linac, a transport directs the beam into a 2 GeV linac extension ($\sim100$ m) which is followed by an $\sim350$~m linac that accelerates the beam to 8 GeV. This is followed by a short transport that injects the 8 GeV beam into an Accumulator Ring. A transfer line ($\sim320$~m long) would transport the accumulated 8 GeV protons into the Recycler Ring at RR-62, or the Main Injector at MI-62. A 2 GeV storage Ring could be added to accumulate beam from the 2 GeV linac for use in low-energy intensity frontier experiments.}
  \label{PIUlinac}
\end{centering}
\end{figure}

Following the Snowmass white paper, Fermilab initiated a Proton Intensity Upgrade (PIU) study to consider potential Fermilab upgrades, including Booster replacement options. In the PIU, an 8~GeV Linac scenario for the Booster replacement was developed, which was similar to that presented above.  A layout for the PIU scenario is presented in Figure~\ref{PIUlinac}.

In the PIU study, design modifications were considered to mitigate risks and provide operational flexibility that could provide beam for physics experiments beyond the 
initial DUNE 2.4~MW requirements. Injection constrained to RR-10 H$^-$ charge exchange injection risks unacceptable irradiation of the MI-10 area, conflicts with MI extraction and restricts linac use for other experiments. Therefore, the scenario includes injection into a new accumulator ring, followed by transfer to the RR or MI in the RR-62 or MI-62 region. RF gradient requirements were reduced from the R\&D goals used above to levels already established. 

\begin{table}[htbp]
\centering
\caption{\label{PIUscenario} Parameters of the PIU 8 GeV linac scenario.}
\smallskip
\begin{tabular}{|l|l|l|l|l|l|}
\hline
\textbf{Section} & \textbf{Length} & \textbf{Bending field} & \textbf{Bending} & \textbf{Cavities/} & \textbf{Cryomodule}\\
 & & \textbf{or} & \textbf{angle or} & \textbf{magnets/} & \textbf{length} \\
& & \textbf{RF frequency} & \textbf{Linac mode} & \textbf{cryomodules} & \\
\hline
1 GeV transport & 75m & 0.277 T & $-45^\circ$ & & \\
1 to 2 GeV linac & 100 m & 650 MHz & CW & 54 / 9 / 9 & 9.92 m \\
2 to 8 GeV linac & 350 m & 1300 MHz & Pulsed, & 208 / 26 / 26 & 12.5 m \\
 & & & 10 -- 20 Hz & & \\
8 GeV Accumulator  & 500 m &  1 T&  $360^\circ$  & & \\
8 GeV injection & 320 m &  & & & \\
\hline
\end{tabular}
\end{table}

\begin{figure}[htp]
\begin{centering}
\includegraphics[width=0.7\textwidth]{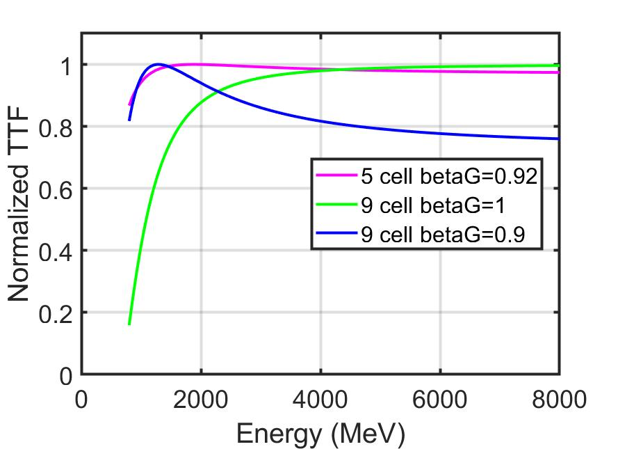}
 \caption{Normalized transit time factors for a $\beta_G=0.92$ 650 MHz 5-cell cavity (as in PIP-II and the 2 GeV extension), a $\beta_G=1$ 1.3 GHz 9-cell cavity (as in European XFEL or LCLS-II), and a potential $\beta_G=0.9$ 1.3 GHz 9-cell cavity.}
  \label{transittimefactor}
\end{centering}
\end{figure}

\begin{figure}[htp]
\begin{centering}
\includegraphics[width=0.7\textwidth]{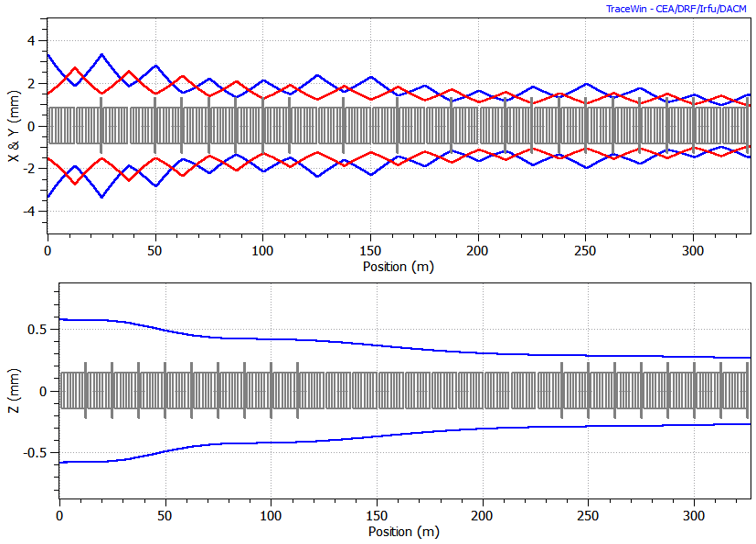}
\caption{Optics for the 2 to 8 GeV linac based on a 31.5 MV/m accelerating gradient. Simulated x, y, and z beam sizes are shown.}
  \label{linacoptics}
\end{centering}
\end{figure}

Table \ref{PIUscenario} summarizes the components of the PIU 8 GeV acceleration scenario. The 800 GeV PIP-II linac is increased to 1 GeV by  the addition of 2 650 MHz cryomodules, placed at the end of the 800 MeV linac. An arc transport takes that beam into a 1 to 2 GeV linac consisting of 9  650~MHz cryomodules. This is followed by a $\sim350$~m 2 to 8 GeV linac using 26 1300 MHz cryomodules. Each cryomodule would contain 8 9-cell $\ \beta_G = 1$ SRF cavities capable of 31.5~MV/m. The normalized transit time factor for 1.3 GHz cavities starts at $\sim0.9$ at 2 GeV (though it quickly improves to $\sim1.0$, as seen in Fig.~\ref{transittimefactor}) and the optics call for the cavities to operate with a synchronous phase of $-10^\circ$ (see Fig.~\ref{linacoptics}).

The 8 GeV linac beam pulses  are accumulated in the 8 GeV storage ring, using H$^-$ stripping or laser-assisted injection, and then transferred at 10 or 20 Hz into the Recycler Ring or Main Injector. The 2 GeV linac would be capable of CW operation and beam from that could be accumulated in a 2 GeV storage ring for use in intensity frontier experiments.

\subsection{Three options for operation of the 2 to 8 GeV linac}

There are different ways to reach 2.4 MW of protons on target for LBNF/DUNE with the linac. We consider 3 options in this scenario: one basic option and two options that deliver significantly more power  in 8 GeV protons,  by increasing the current from PIP-II and/or by increasing the RF duty factor. The three options are summarized below:

\begin{itemize}
\item A -- Basic option: slight PIP-II current upgrade, uses European XFEL RF sources
\item B -- High current PIP-II option: significant PIP-II current upgrade, requires some RF R\&D
\item C -- Highest power option: significant PIP-II current upgrade, requires significant RF R\&D
\end{itemize}

Option A meets the requirements for LBNF/DUNE without any major R\&D development for the RF sources, but  just using what has been demonstrated with XFEL sources. 
For option A, a beam current of 2.7 mA is specified during the pulses, this may require upgrading the RF sources for the LB650 cryomodules in PIP-II to allow them to reach the  higher peak power.
For Option B, PIP-II should be upgraded to allow higher current operation, and that would require upgrades to its RF sources. For Option C, more significant upgrades are needed  to enable higher duty factor operation (longer pulses, higher repetition rate).

The three options will have impact on the extra beam power available for the 8 GeV physics program.  Option A provides a relatively small amount of extra power for the 8 GeV program, $\sim160$~kW. Options B and C provide significantly more power (more details below). With option C, which combines higher duty factor RF and higher beam current, MW-scale beam power is provided at 8~GeV, which could enable a muon collider or neutrino factory.

Table \ref{tab:viii} presents some important parameters for the RF system in the three options A, B, and C. All parameters were based on a cavity $Q_0$ of 1$\times10^{10}$, except for the bottom line, which shows how the 2~K heat load would change for a $Q_0$ of 2$\times10^{10}$ (requires SRF R\&D). 
The following key assumptions were made for the RF sources for all three options: 1) power overhead of 25\%, 2) maximum average power per RF source is <~150~kW, 3) input power is the same for filling and acceleration (options B and C).

\begin{table}[ht]
\centering
\caption{\label{tab:viii}Summary of RF parameters for PIU  linac scenarios.}
\smallskip
\begin{tabular}{|l|l|l|l|}
 \hline

\textbf{Option}                & \textbf{A}            & \textbf{B}            & \textbf{C}           \\ \hline
 Beam current, mA                                     & 2.7 & 5.0 & 5.0 \\ \hline
Flat-top pulse length, msec                           & 1.50 & 2.0 & 2.0 \\ \hline
Rep. rate, Hz                                         & 10  & 10  & 20  \\ \hline
Optimal loaded $Q$, $\times10^{6}$                                & 4.5        & 4.5        & 4.5       \\ \hline
Filling time, msec                                             & 0.6          & 1.0          & 1.0          \\ \hline
Pulse width, msec                                              & 2.1          & 3.0          & 3.0          \\ \hline
Duty factor, \%                                                & 2.1          & 6.0          & 6.0          \\ \hline
Pulsed input power / cavity, filling, kW                         & 330          & 170          & 170          \\ \hline
Pulsed input power / cavity, acceleration, kW                    & 110          & 170          & 170          \\ \hline
Average input power / cavity, kW                                 & 3.6          & 5          & 10            \\ \hline
RF source pulsed power/cryomodule, MW                                  & 3.3          & 1.7            & 1.7          \\ \hline
RF source average power / cryomodule, kW                                 & 37           & 50           & 100           \\ \hline
No. of cryomodules / RF source (150 kW limit)                            & 3            & 3            & 2            \\ \hline
RF source pulsed power, MW                                     & 10           & 5            & 3.4            \\ \hline
RF source average power, kW                                    & 110          & 150          & 200          \\ \hline
Cryogenic load for baseline $Q_0$ of 1$\times10^{10}$, W & 617          & 754         & 1360          \\ \hline
Cryogenic load for $Q_0$ of $2\times10^{10}$, W & 390          & 456          & 762     \\  \hline
 \end{tabular}

\end{table}


Option A is based on the capabilities of the European XFEL multi-beam klystron RF system. Options B and C use a pulse width of 3.0 msec, which is beyond the capability of the  XFEL RF system; these options require RF source R\&D. Option C parameters are roughly based on the higher average power, lower pulsed power SBK CPI VKL7796 klystron, but further R\&D could be done to improve the parameters. To accommodate the higher average RF power, some modifications should also be done for the input coupler. Designs and tested prototypes already exist for a high power 1.3~GHz coupler \cite{Kazakov2018,Veshcherevich2003}.

For option B and C, PIP-II CMs would operate with CW RF up through the LB650s. The HB650s (for  both PIP-II and the 2 GeV extension) would  operate pulsed with 5 mA beam current and 20 Hz rep rate, with half of the beam current going to the 8 GeV linac (operated at 10 Hz), and half going to a 2 GeV program (10 Hz to this program).
For option C, 20 Hz pulses are sent through the 2 to 8 GeV linac to deliver substantially higher 8 GeV beam power. RF source work should be done to develop sources that reach the specifications, similar to has been done in the past for previous accelerators.

Table \ref{linacbeam} presents some important parameters for the beams from the 8 GeV linac in the three scenarios. All three get 2.4 MW beam power for LBNF/DUNE. For Option B and C, there are also configurations that successfully reach 2.4 MW for LBNF/DUNE without use of the Recycler to accumulate batches. For Option A, the Recycler is needed to achieve 2.4 MW of beam power to LBNF/DUNE.

\begin{table}[ht]
\centering
\caption{Summary of parameters for the beams from the 8 GeV linac.}
\label{linacbeam}
\smallskip
\begin{tabular}{|l|l|l|l|}
\hline

\textbf{Option}                & \textbf{A}            & \textbf{B}            & \textbf{C}   \\ \hline
Beam current, mA          & 2.7  & 5  & 5  \\ \hline
Flat-top pulse length, msec & 1.5 & 2   & 2  \\ \hline
Rep. rate, Hz               & 10  & 10  & 20 \\ \hline
Max pulse charge, $10^{-6}$ C            & 4.05         & 10          & 10          \\ \hline
Number of batches                    & 6            & 6            & 6           \\ \hline
Batches / pulse                      & 1            & 2            & 1           \\ \hline
Charge / batch, $10^{-6}$ C               & 4            & 4            & 4           \\ \hline
Required pulse at current, ms        & 1.49         & 1.62         & 0.8        \\ \hline
Required proton / batch, $\times10^{12}$       & 25           & 25           & 25          \\ \hline
Required MI intensity, $\times10^{12}$         & 150          & 150          & 150         \\ \hline
Accumulation time, s                 & 0.5          & 0.2         & 0.25         \\ \hline
MI cycle time, s                     & 1.2          & 1.2          & 1.2         \\ \hline
MI power, MW                         & 2.4         & 2.4         & 2.4        \\ \hline
Available cycles for 8 GeV           & 50\%       & 75\%       & 75\%      \\ \hline
Available power at 8 GeV, kW         & 162          & 570          & 1200         \\ \hline
Potential MI power, MW            & 2.4         & 2.8         & 3.2       \\ \hline
\end{tabular}

\end{table}

In Options B and C, the 2.0 ms linac pulse delivers more than what is necessary to fill the MI over 6 batches. Accordingly, the minimum pulse length needed to fill at that current is given. But for the 8 GeV program, we assume the full 2.0 ms pulse is delivered and that the AR can handle that charge. Options A, B, C  have significant variations in 8 GeV beam power, owing first to the difference in linac current and next to the difference in the linac pulse rate.  The potential MI power is the beam power at 120 GeV, if we limit the Main Injector to $200 \times 10^{12}$ protons but set aside power limits on DUNE/LBNF beamline.

The beam power associated with option C overlaps with the beam power requirements of the Muon Collider proton driver. To support this possibility,
 higher linac beam current and/or duty factor could be increased even further.

\section{R\&D requirements}

The Booster Replacement program will require a significant R\&D program to obtain a timely implementation of the required upgrade. The DOE project process requires detailed designs and evaluations of the proposal and alternatives, which implies that both RCS and linac-based approaches should be evaluated and compared.
The SRF linac is based on the PIP-II cryomodules for the 650~MHz part and ILC/LCLS-II for the 1300~MHz part. The designs should be updated to incorporate recent improvements in SRF and optimized for cost and efficiency.

Other R\&D topics that need to be investigated include:
\begin{itemize}
    \item Simulation and modelling of the complete linac design, from PIP-II into the MI.
    \item Simulation and optimization of the injection painting and foil heating.
    \item Consideration of laser-assisted injection and its adaptation to the BRL and RCS scenarios.
    \item An evaluation of SRF power and wall-plug power requirements for the scenarios, for pulsed and CW operation options, including optimizations, should be developed. 
    \item An alternative injection into a new $\sim 8$ GeV storage ring could be considered; this would avoid the MI-10 bottleneck, but at the cost of an additional storage ring. The ring may be needed for intensity frontier experiments.
\end{itemize}

Some details of the proposed R\&D program are outlined below.

\subsection{SRF research topics}

Here is a list of key SRF research topics that will be part of the BRL R\&D program:
\begin{itemize}
    \item 650 MHz: higher $Q$ $(>6\times10^{10})$ at 2 K and 20.9 MV/m -- using either improved nitrogen doping or medium temperature baking.
    \item 1300 MHz: high $Q$ of $2\times10^{10}$ at 2 K and higher gradient of $\simeq 33.7$~MV/m -- using a new 2-step low temperature bake or some other surface treatment recipe.
    \item Resonance control R\&D for microphonics suppression in CW mode and Lorentz Force Detuning (LFD) compensation in pulsed mode for cavities operating at high loaded $Q$ factors.
    \item Ferroelectric tuner for both resonance control and coupling adjustment -– will improve efficiency of the SRF systems.
    \item Robotic assembly of the SRF cavity strings in clean rooms –- essential for achieving high gradients.
\end{itemize}

Most of these R\&D topics are in line with the \textit{DOE HEP General Accelerator R\&D RF Research Roadmap} \cite{25}.

\subsubsection{Improving the 650 MHz cavity performance}

\begin{figure}[htbp]
\centering 
\includegraphics[width=.8\textwidth]{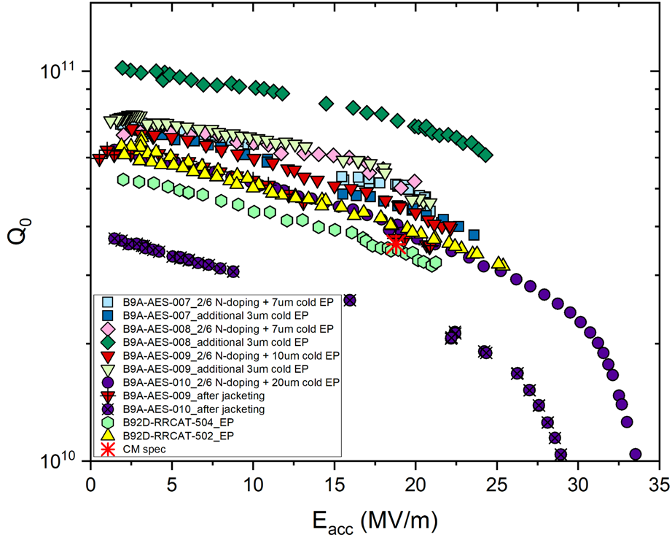}
\qquad
\caption{\label{fig:17} Vertical acceptance test of HB650 $\beta_G$ = 0.9 cavities \cite{28}.}
\end{figure}

Recent progress in SRF experimental and theoretical research resulted in a dramatic increase in achievable quality factors by a) the development of new surface treatments resulting in very high $Q$ via nitrogen doping \cite{5}; and b) achievement of very high $Q$ under real accelerator conditions via efficient magnetic flux expulsion (fast cooling and low flux pinning) \cite{6,26}. These advances have been confirmed at laboratories worldwide and transferred to industry. They have found practical demonstration in the LCLS-II cryomodules that have reached two times the previous state of the art $Q$ on an accelerator scale unit with an average of $Q \approx 3\times10^{10}$ at 2 K, 1.3 GHz, 16 MV/m \cite{27}. Further improvement of the nitrogen doping recipe allows the LCLS-II-HE project to increase the accelerating gradient to 20.8 MV/m while maintaining high quality factors of SRF cavities. PIP-II has adapted nitrogen doping to their 5-cell 650 MHz elliptical cavities, with HB650 cavities ($\beta_G$ = 0.92) operating at 18.8 MV/m with $Q > 3\times10^{10}$ \cite{28}. Fig.~\ref{fig:17} shows results of testing four cavities in the vertical test cryostat.

There are still ample R\&D opportunities for improving the quality factor of 650 MHz SRF cavities to achieve the parameters desired for the BRL: 20.9 MV/m with $Q > 6\times10^{10}$. As we can see in Fig.~\ref{fig:17}, obtaining the gradient is already within reach. However, dedicated R\&D on improving the nitrogen doping recipe will be needed to reliably achieve the required quality factor.

\subsubsection{High gradient / high Q R\&D at 1300 MHz}

\begin{figure}[htbp]
\centering 
\includegraphics[width=.8\textwidth]{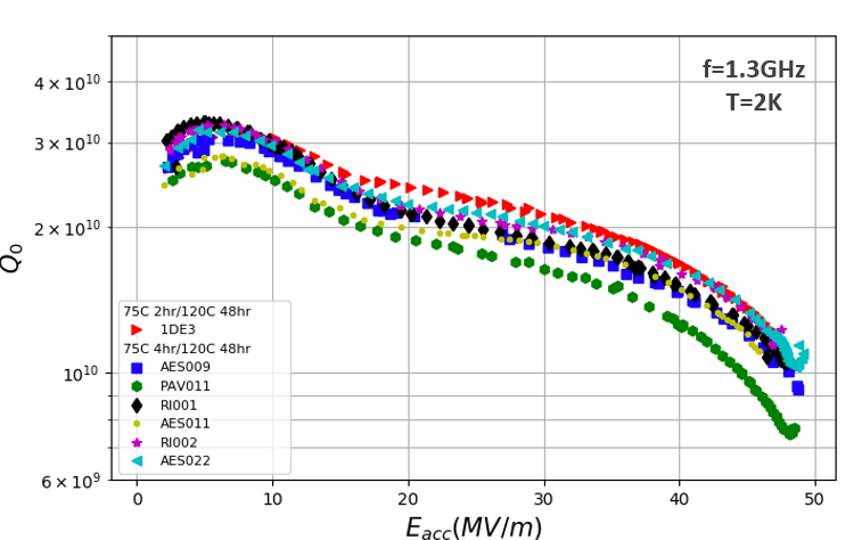}
\qquad
\caption{\label{fig:18} Vertical tests of single cell TESLA cavities treated with the new recipe.}
\end{figure}

For the 1300 MHz pulsed linac, we will need a surface treatment recipe that would reliably produce cavities with the accelerating gradient of 33.7 MV/m and $Q > 2\times10^{10}$. Recently, as part of the ILC cost reduction R\&D, a new treatment has been developed that combines ultra-cold electropolishing and a two-step low temperature bake \cite{29}. Cavities treated with this recipe systematically achieve extremely high gradients of $\sim 48$ -- 50 MV/m but at lower Q than we require for the BRL, see Fig.~\ref{fig:18}. Further studies must be performed to optimize the new recipe for BRL needs.

\subsection{Resonance control of SRF cavities}

R\&D for high acceleration gradient and high $Q$ should be accompanied by cavity resonance control development \cite{30,31}. SRF structures operating in the CW regime, like the 650 MHz linac for the BRL, are susceptible to vibrations due to external excitation (microphonics). For pulsed-beam accelerators such as the 1300 MHz pulsed linac for BRL, compensating cavity resonant frequency detuning due to the Lorentz force is especially important as the ratio of LFD over the cavity bandwidth is proportional to the cube of the acceleration gradient. The sources of microphonics should be determined, understood, and mitigated. A robust active LFD correction with microphonics compensation should be developed for the BRL. The correction algorithm development shall be done in conjunction with developing fast cavity frequency tuners, based on either the currently dominant piezo-electric mechanical tuner technology or on a new ferroelectric technology described below.

\subsection{Ferroelectric tuner R\&D}

\begin{figure}[htbp]
\centering 
\includegraphics[width=.6\textwidth]{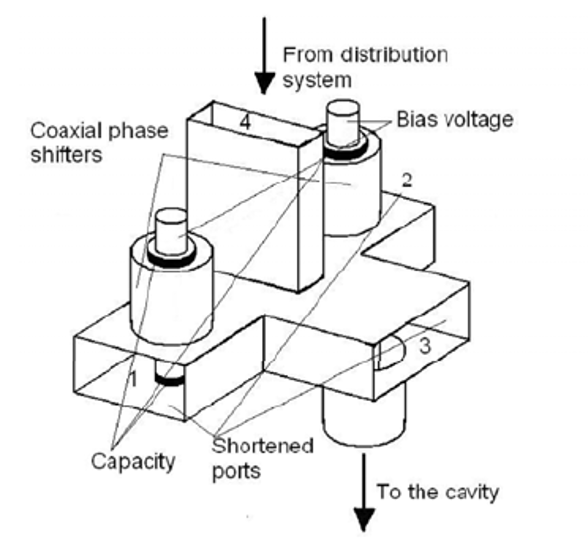}
\qquad
\caption{\label{fig:19} Schematic of a device to produce fast cavity coupling changes based on a magic-T and two phase shifters containing ferroelectric elements. This device is not under vacuum and is located outside of the cryostat.}
\end{figure}

Recently developed ferroelectric ceramics with low losses at RF frequencies \cite{32,33,34} allow the development of electrically controlled tuners with switching times much better than that of piezoelectric mechanical tuners. Such a tuner is inserted into a high-power transmission line connected to the SRF cavity and allows alteration of coupling between the acceleration structure and the line during the RF pulse \cite{35}. In addition, this tuner would allow electronic control of the cavity frequency by a device operating at room temperature within timescales of active compensation of microphonics. Thus, a ferroelectric tuner can perform the double function of coupling and frequency tuning. It would reduce cryogenic losses in the structure and consequently significantly reduce the overall energy consumption of the accelerator and could eliminate the need to use over-coupled fundamental power couplers, thus significantly reducing RF amplifier power. A proof-of-principle demonstration of the ferroelectric fast reactive tuner (FE-FRT) was successfully conducted at CERN \cite{36}. The timescale in which the FE-FRT is able to shift the cavity frequency across the entire tuning range was measured to be < 50 $\mu$s; this is significantly faster than any other cavity tuning device. The measurement was limited by the signal to noise ratio and the true timescale could well be more than an order of magnitude lower still. The experiment paves a way toward developing a device that can produce fast cavity coupling change, such as shown in Fig.~\ref{fig:19}. More R\&D efforts are needed in this area to fully realize the full potential of ferroelectric devices.

\subsection{Robotic assembly of SRF cavity strings}

Field emission phenomena can be a serious impediment to achieving high gradients. Special studies will be required, in parallel with the high gradient research for the BRL, to abate field emission in SRF cryomodules. One of the promising pathways is robotic assembly of SRF cavities and systems in clean rooms. Using robots for automated assembly would eliminate cavity contamination and assembly inconsistencies due to the ``human factor'', which is the dominant cause of field emission cavity performance degradation. This is a nascent R\&D area that is just starting to receive proper attention at various laboratories around the world.

\subsection{Laser-assisted injection R\&D}

Foil heating and damage, as well as beam losses associated with foil-based injection, are a significant limitation on the performance of high-intensity H$^-$ injected beams, including both the RCS and the linac-based versions of the Booster Replacement. Research has demonstrated that laser-assisted injection can be used for a proton ring \cite{37,41}.  The 1 GeV proton energy of that demonstration required relatively difficult ultraviolet light, which limited the pulse length.
Laser-assisted injection is considered to be the eventual preferred procedure for H$^-$ injection, but the R\&D needed for implementation has not yet been performed. Laser stripping of high-energy hydrogen atoms is relatively easy, because the laser light is Doppler-shifted to higher energy in the rest frame following

\begin{equation}
\lambda_{PF} = \frac{\lambda_{Lab}}{\gamma (1+\beta \cos\theta)},
\end{equation}
where $\gamma$ = 9.526 and $\beta$ = 0.9945 for 8 GeV protons, and $\theta$ is the angle between the laser and particle beam, which should be nearly collinear ($\theta \cong 0^\circ$). The laser intensity $I_0$ in the lab frame is boosted in the beam frame to

\begin{equation}
I_{PF} = I_0 (1+\beta \cos\theta)^2 \gamma^2.
\end{equation}

\begin{figure}[htbp]
\centering 
\includegraphics[width=1.0\textwidth]{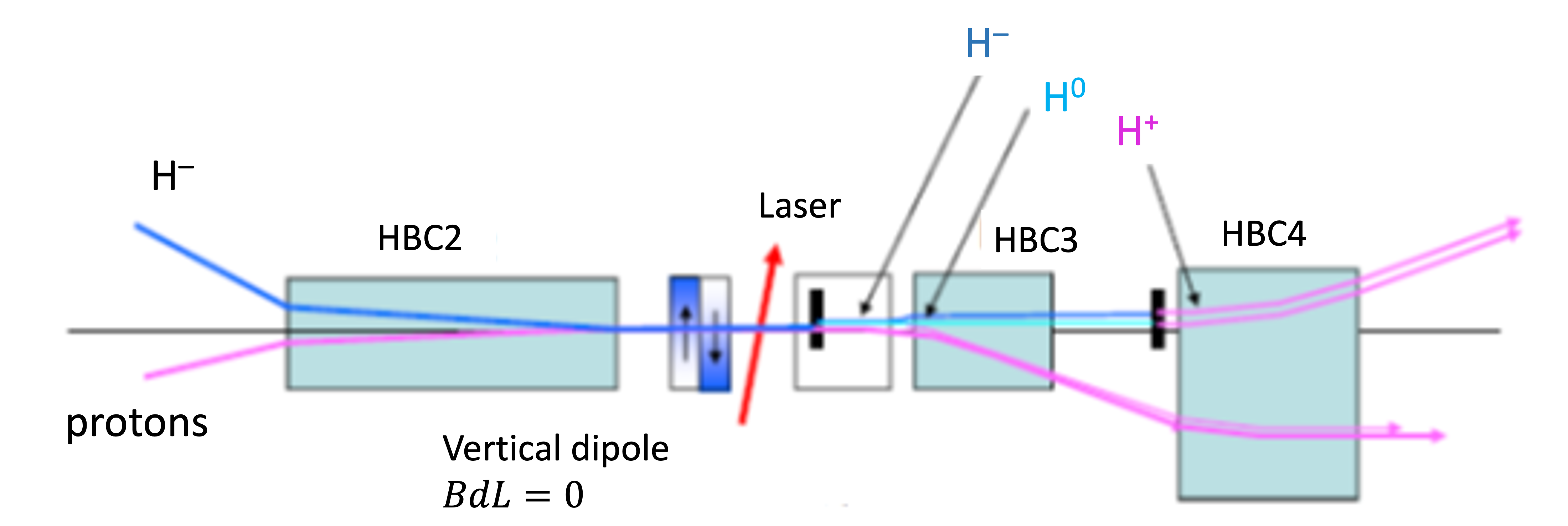}
\qquad
\caption{\label{fig:20} Schematic view of a laser-stripping injection system that could replace foil injection. The new insert consists of a vertical dipole with peak field of 3~kG and $BdL = 0$, which would have minimal effect on circulating beam and strip the outer electron from H$^-$, and a laser crossing that would replace foil stripping.}
\end{figure}

Hydrogen ionization requires $\sim90$ nm photons, which can be obtained from Doppler-shifted $\sim1700$~nm laser light, and the intensity would be magnified by a factor of 360. A high intensity infrared (1000 – 1700 nm) laser matched to the proton injection pulse could be possible \cite{38}. Fig.~\ref{fig:20} shows a simplified view of a possible stripping configuration. A redesign of the injection scenario to include laser injection should be developed and compared to the baseline foil injection. This should be followed by the acquisition and testing of some of the hardware needed for laser injection, possibly including a high-power laser. A beam test at Fermilab may be possible.

\acknowledgments

This work was produced by Fermi Research Alliance, LLC under contract No. DEAC02-07CH11359 with the U.S. Department of Energy. The United States Government retains and the publisher, by accepting the work for publication, acknowledges that the United States Government retains a non-exclusive, paid-up, irrevocable, world-wide license to publish or reproduce the published form of this work, or allow others to do so, for United States Government purposes. The Department of Energy will provide public access to these results of federally sponsored research in accordance with the DOE Public Access Plan (http://energy.gov/downloads/doe-public-access-plan).

\end{document}